\documentclass[10pt,a4paper]{article}
\usepackage{jheppub}
\usepackage[utf8]{inputenc}
\usepackage[T1]{fontenc}
\usepackage{amsmath}
\usepackage{amssymb}
\usepackage{graphicx}
\usepackage{MnSymbol}
\usepackage{comment}
\usepackage{multirow}
\usepackage{booktabs}

\newcommand{\mE}{\mathbb E}

\newcommand{\bb}[1]{\boldsymbol{#1}}
\newcommand{\bbb}[1]{\bar{\boldsymbol{#1}}}
\renewcommand{\det}{\ensuremath{\text{det}}}

\title{Fundamental operating regimes, hyper-parameter fine-tuning and glassiness: towards an interpretable replica-theory for trained restricted Boltzmann machines}

\author[a]{Alberto Fachechi,}
\author[a]{Elena Agliari,}
\author[a]{Miriam Aquaro,}
\author[b,c]{Anthony Coolen,}
\author[b]{Menno Mulder}

\affiliation[a]{Dipartimento di Matematica, Sapienza University of Roma, 
P. le A. Moro 5, 00185 Roma, Italy}
\affiliation[b]{Theoretical Biophysics, DCN Donders Institute, Faculty of Science,
Radboud University, 6525 AJ Nijmegen, The Netherlands}
\affiliation[c]{Saddle Point Science Europe, Mercator Science Park, 6525 EC Nijmegen, The Netherlands}
\emailAdd{alberto.fachechi@uniroma1.it}

\abstract{
We consider restricted Boltzmann machines with a binary visible layer and a Gaussian hidden layer trained by an unlabelled dataset composed of noisy realizations of a single ground pattern. We develop a statistical mechanics framework to describe the network generative capabilities, by exploiting the replica trick and assuming self-averaging of the underlying order parameters (i.e., replica symmetry). In particular, we outline the effective control parameters (e.g., the relative number of weights to be trained, the regularization parameter), whose tuning can yield qualitatively-different operative regimes. Further, we provide analytical and numerical evidence for the existence of a sub-region in the space of the hyperparameters where replica-symmetry breaking occurs.}

\begin{document}
     \date{\today}
%\graphicspath{{./Figures}}

\maketitle
 \tableofcontents

\newpage

\section{Introduction and related work}
Boltzmann Machines (BMs) \cite{ackley1985learning,smolensky1986information} are paradigmatic energy-based models in Machine Learning, mainly employed for estimating an unknown probability distribution underlying a given set of data in an unsupervised way. More specifically, the model probability distribution can be looked at as the Boltzmann-Gibbs distribution of a spin system, with the weights and the biases playing the role of, respectively, couplings and external magnetic fields. The training equations can be obtained by minimizing the Kullback-Leibler (KL) divergence between the model distribution and some target distribution and exhibit the structure of moment-matching criteria \cite{salakhutdinov2008learning,hinton2010practical}. Countless applications of BMs (together with higher-order \cite{sejnowski1986higher} as well as deep \cite{salakhutdinov2009deep} extensions) have been analyzed in the literature, see for instance \cite{prager1986boltzmann,larochelle2008classification,tang2012robust,larochelle2012learning,srivastava2013modeling,hu2014deep,torlai2016learning,carleo2017solving,nomura2017restricted,melko2019restricted,deep2021,yelmen2021creating,mauri2023mutational,sekimoto2024quasi,fernandez2023designing} just to mention a few. Because of the generality of the problem under consideration and the possibility to develop a grounded mathematical theory for such models, BMs have attracted significant theoretical interest in the recent years as minimal models for inferential and representation capabilities \cite{jebara2012machine,tubiana2017emergence,decelle2021exact,decelle2021inverse,decelle2023unsupervised,fernandez2023disentangling,decelle2024inferring} as well as generative functionalities \cite{ventura2023unlearning,roussel2023accelerated,roussel2023deep,roussel2021barriers,yasuda2024effective}. In fact, a large variety of possible approaches to their analysis have been developed, with the final goal of describing quantitatively the effectiveness of the learning dynamics \cite{gabrie2015training,huang2015advanced,takahashi2016mean,decelle2017spectral,decelle2018thermodynamics,ccakmak2020dynamical,decelle2021exact,decelle2021equilibrium} and the related performances in the operative regime after training \cite{barra2017phase,barra2018phase,agliari2020neural,agliari2021learning}. These theoretical investigations have highlighted important drawbacks in a straightforward usage of BMs in practical situations, mainly due to the fact that training effectiveness depends on the architecture and on the value of the learning hyperparameters, and that the resulting inferred models can be close to critical points \cite{mastromatteo2011criticality,tkavcik2015thermodynamics}, thus harming the capability of generating new samples from the network due to long-temporal correlation in the Monte Carlo Markov Chain, see also \cite{decelle2021equilibrium}. Thus, several {\it ad hoc} regularization techniques \cite{ventura2023unlearning}, suitable weight initialization, pre-training programs \cite{hinton2012better,pozas2021efficient,leonelli2021effective} or mean-field-based effective training procedures \cite{kappen1998efficient,cho2011improved} have to be carried out. Despite all this progress, a solid grounded theory describing the working regimes of BMs (and thus also providing hyperparameters fine-tuning prescriptions) is still under construction, see also \cite{fernandez2024replica,bachtis2024cascade} for recent steps in this direction. A crucial point in this program is the deep relation with spin-glass models  \cite{barra2012equivalence,fachechi2019dreaming,marullo2020boltzmann,decelle2021restricted,Huang2021}, a bridge allowing the usage of statistical-mechanical techniques to address the relaxation towards equilibrium \cite{mezard1987spin,fernandez2024replica,hoyle2023replica}, and possibly to derive rigorous results \cite{agliari2019free,genovese2020legendre,alberici2020annealing}. The key point lies in the fact that training procedure can be conceptualized as a dynamical process assigning (in a Bayesian point of view) the posterior probability distribution of the weights conditioned on the empirical data. In this scenario, the weights can be treated as the fundamental degrees of freedom, while the dataset plays the role of the quenched disorder. Such an approach can be for instance used to derive effective theories describing kernel renormalization in deep feed-forward and convolutional neural networks, see for example \cite{li2021statistical,pacelli2023statistical,aiudi2023local}. 
\par\medskip
In this paper, we consider a binary-Gaussian Restricted Boltzmann Machine ({\it vide infra}) as a model to be trained by extracting information from a dataset characterized by a single {hidden} ground-truth. In particular, we derive a statistical mechanics picture for the network in the generative regime, also giving a complete characterization of the solution under the replica-symmetry (RS) assumption. Despite the simplicity of the setting under consideration, the resulting phenomenology turns out to be very rich, exhibiting in particular existence conditions for the RS solution in the space of hyperparameters (thus resulting in a fine-tuning prescription for the setting under consideration). Numerical investigations show evidence for ageing behavior of the system outside the RS region, and strongly suggest the intrinsic replica-symmetry breaking (RSB) behavior for such choices of the hyperparameters. The paper is organized as follows. In Sec.~\ref{sec:RBM} we introduce the model under investigation and we derive the related learning rules. Next, in Sec.~\ref{sec:replica}, we specify the statistical properties of the datasets considered here and we carry our the analytical investigation, highlighting the existence of different working regimes as the value of the hyperparameters is tuned. Then, in Sec.~\ref{sec:selfcons}, we progress the investigation by relying on numerical methods and, in Sec.~\ref{sec:numerical}, the results collected are corroborated by Monte Carlo simulations. Finally, Sec.~\ref{sec:outro} contains our conclusions and an outlook. The appendices are dedicated to technicalities and details on calculations.
 
\section{Restricted Boltzmann Machines}\label{sec:RBM}
Restricted Boltzmann Machines (RBMs) are energy-based models which can be seen as bipartite spin-glass models, with interacting neurons $\{\sigma_i\}_{i=1}^{N_\sigma}$ constituting the visible layer and $\{s_\mu\}_{\mu=1}^{N_s}$ making up the hidden one, whose distribution is governed by the Boltzmann-Gibbs measure
\begin{equation}
	P_{\bb W}(\bb\sigma,\bb s )=\frac 1{Z_{\bb W}} \exp\big(-\beta_1 H_{\bb W}(\bb \sigma,\bb s )\big),
\end{equation}
with $\beta_1$ being the inverse sampling temperature, $Z_{\bb W}$ the partition function of the model, and 
\begin{equation}\label{eq:H}
	\begin{split}
		H_{\bb W}(\bb \sigma,\bb s)=& -\sum_{i\mu }W_{i\mu}\sigma_i s_\mu,
	\end{split}
\end{equation}
is its energy function.\footnote{In this paper, we examine the case without biases. Their inclusion however does not change in a substantial way the methodological scheme developed here.} BMs are traditionally used as feature-extractor generative models, as they can be used to capture information about the statistics underlying a given dataset. Let $\mathcal D=\{\bar{\bb\sigma}^A\}_{A=1}^M$ be a set of data sampled from an unknown probability distribution $Q(\bb\sigma)$ that we want to model by means of the marginal distribution $P_{\bb W}(\bb\sigma)= \sum_{s} P_{\bb W}(\bb\sigma,\bb s )$. The BM’s training is accomplished by minimizing a suitable objective function that quantifies the dissimilarity between the probability distributions, usually the KL divergence:
\begin{equation}\label{eq:KL}
	D_{KL} (P_{\bb W}\lVert Q)=\sum_{\bb\sigma} Q(\bb\sigma)\log\frac{Q(\bb\sigma)}{P_{\bb W}(\bb\sigma)}.
\end{equation} 
Since we have no direct access to the distribution $Q(\bb\sigma)$, but only indirectly by means of a (quenched) realization of the dataset $\mathcal D$, one has to consider the empirical version of \eqref{eq:KL}, thus replacing $\sum_{\bb\sigma} Q(\bb\sigma) f(\bb\sigma)\to \frac1M\sum_{\bar{\bb\sigma}\in \mathcal D}\delta_{\bb\sigma, \bar{\bb\sigma}} f(\bb\sigma)$. {We will denote as $\hat D _{KL}$ the empirical KL divergence. The minimization can be achieved by the gradient descent algorithm}
\begin{eqnarray}\label{eq:det_GD}
	\frac{d \bb W}{dt}=-\nabla_{\bb W}{\hat D}_{KL}(P_{\bb W}\lVert Q),
\end{eqnarray}
so that fixed points realize the extremality condition $\nabla_{\bb W}\hat D_{KL}=0$. {The empirical KL divergence can be further simplified as
\begin{equation}
	\begin{split}
		{\hat D}_{KL}(P_{\bb W}\lVert Q)=-\frac1M \sum_{A=1}^M\log\sum_{\bb s} P_{\bb W}(\bar{\bb\sigma}^A)-\hat {H} (Q),
	\end{split}
\end{equation}
where $\hat H(Q)$ is the empirical entropy of $Q$, which is independent of the network parameters, and therefore gives rise to negligible contribution in the training equations \eqref{eq:det_GD}.} Straightforward computations show that the gradient flow dynamics can be expressed in terms of relevant correlation functions as
\begin{eqnarray}
\label{eq:CD}
	\frac{d W_{i\mu}}{dt}&=& \beta_1 (\langle \sigma_i s_\mu\rangle_{\bb W,+}-\langle \sigma_i s_\mu \rangle_{\bb W,-}),
\end{eqnarray}
where the subscripts ``$\bb W,+$'' and ``$\bb W,-$'' distinguish between clamped and free average, respectively. In fact, the correlation functions (also referred to as resp. positive and negative gradients) appearing in the previous equations are defined as specified hereafter for a generic observable $O(\bb\sigma,\bb s)$:
\begin{eqnarray}
\langle O(\bb\sigma,\bb s) \rangle_{\bb W,+} &=& \frac1M \sum_{A=1}^M  \sum_{\bb s} O(\bar{\bb\sigma}^A,\bb s) P _{\bb W}(\bb s \vert \bar{\bb \sigma}^A),\\
\langle O(\bb\sigma,\bb s) \rangle_{\bb W,-} &=&  \sum_{\bb\sigma,\bb s } O(\bb\sigma,\bb s) P  _{\bb W}(\bb \sigma,\bb s).
\end{eqnarray}
Eq. \eqref{eq:CD} is commonly referred to as moment-matching criterion, since the fixed point of the gradient-flow equation yields a match between the 2-point correlation function w.r.t., respectively, the conditioned probability of the model with the visible layer clamped on the input data and the joint probability distribution. Naively speaking, in this way the model captures the first two moments of the statistics $Q(\bb\sigma)$ generating the dataset $\mathcal D$.

\section{Replica computations for binary-Gaussian Restricted Boltzmann Machines}\label{sec:replica}
In this section, we enlarge the setting described previously in order to reach a statistical-mechanics framework where neurons $\bb \sigma,\bb s$ and learning parameters $\bb W$ are both meant as degrees of freedom.
The starting point is to modify the evolution equation \eqref{eq:CD} in order to give it the form of a stochastic process (thus mimicking the stochastic version of gradient flow dynamics). This can be achieved by adding a damping factor (i.e., an $L_2$-regularization or, in a Bayesian jargon, setting a Gaussian prior on the weights $\bb W$) and a noisy contribution $\boldsymbol{\eta}$, so that
\begin{eqnarray}\label{eq:langevin}
	\frac{d W_{i\mu}}{dt }&=& -\frac{\partial \mathcal H(\bb W)}{\partial W_{i\mu}}+\eta_{i\mu}(t),\label{eq:stochastic1}
\end{eqnarray}
where
\begin{eqnarray}\label{eq:H_W}
\mathcal H (\bb W) =\hat  D_{KL}(P_{\bb W}\lVert Q)+\frac{\epsilon_WN }{2}\sum_{i\mu}W_{i\mu}^2,
\end{eqnarray}
is the objective function, $\epsilon_W$ is the regularization hyperparameter, and
$$
\mE_{\eta} \eta_{i\mu}(t) = 0,\quad \mE_{\eta} \eta_{i\mu}(t)\eta_{j\nu}(t')=2T_2 \delta_{ij}\delta_{\mu\nu}\delta(t-t'),
$$
with $T_2 \in \mathbb R^+$ tuning the variance of the noise {and $\mathbb E_{\eta}$ denoting the standard Gaussian mean over the i.i.d. random variables $\eta_{i\mu}$}. In a physical context, Eq. \eqref{eq:langevin} is a Langevin-type stochastic differential equation with objective function $\mathcal H$, with $T_2$ playing the role of a temperature and here will be therefore referred to as {\it learning temperature}; remarkably $T_2$ can be linked to the mini-batch size in the stochastic gradient descent approach, see App. \ref{app:T2} for a discussion. From the theory of stochastic processes, it is well-known that the equilibrium dynamics of the Langevin equation is described by a Boltzmann-Gibbs distribution $\mathcal P (\bb W)={\mathcal Z}^{-1}\exp\big(-\beta_2 \mathcal H(\bb W)\big),$ with $\beta_2=T_2^{-1}$, and
\begin{equation}
	\begin{split}
		\mathcal Z& %= \int d\bb W  d \bb\theta \exp(-\beta_2 \mathcal H (\bb W,\bb\theta) )=
		%\\&
		=\int d\bb W \exp\Big(\frac{\beta_2}M \sum_{A=1}^M \log \sum_{\bb s} P_{\bb W}(\bar{\bb\sigma}^A,\bb s )-\frac{\beta_2 \epsilon_WN}{2}\sum_{i\mu} W_{i\mu}^2\Big),
	\end{split}
\end{equation}
is the partition function. The distribution $\mathcal P (\bb W)$ thus plays the role of posterior distribution of the weights. The first contribution in the exponential can be rewritten as
\begin{equation}
	\begin{split}
		\frac{\beta_2}M \sum_{A=1}^M \log \sum_{\bb s}P_{\bb W} (\bar{\bb \sigma}^A,\bb s)
		= \beta_2 \Big(\overline{\log Z_{\bb W}(\bar{\bb \sigma}^A)}-\log Z_{\bb W}\Big),
	\end{split}
\end{equation}
where $\overline{\ \cdot\ }$ stands for the empirical mean on the dataset, and $Z_{\bb W}(\bar {\bb \sigma}^A)= \sum_{\bb s} \exp(-\beta H_{\bb W}(\bar{\bb\sigma}^A,\bb s))$ is the partition function where the visible layer is clamped to a given example $\bar{\bb\sigma}^A$ in the dataset. Thus, we can recast the partition function as
\begin{equation}
	\begin{split}
	\mathcal Z& =\int d\bb W \exp\big(-\frac{\beta_2N\epsilon_W}{2}\sum_{i\mu} W_{i\mu}^2\big)  Z_{\bb W}^{-\beta_2}\prod_{A=1}^M  Z_{\bb W}(\bar{\bb\sigma}^A)^{\frac{\beta_2}M}.
	\end{split}
\end{equation}
The free energy per site associated to the partition function $\mathcal Z$ can be computed as $\mathcal F (\mathcal D)= -\frac{1}{\beta_2 N} \log \mathcal Z$ for a given realization of the dataset $\mathcal D$, {where $N=N_\sigma+N_s$}. Performing the quenched average over all possible realizations of the dataset with same statistics $Q$, and taking the thermodynamic limit ${N\to\infty}$ such that $N_s/N_\sigma=\alpha\in \mathbb R_+$, we can write
\begin{equation}\label{eq:tot_fe}
	\begin{split}
	\bar{ \mathcal F}_M %&=\lim_{N\to\infty}-\frac1{\beta_2N}\log\int d\bb W  d \bb\theta \exp\Big(-\frac{\beta_2N\epsilon_W}{2}\sum_{x<y} W_{xy}^2-\frac{\beta_2\epsilon_\theta}{2}\sum_{x} \theta_x^2\Big) Z(\bb W,\bb\theta)^{-\beta_2}\prod_{A=1}^M Z(\bar{\bb\sigma}^A,\bb W,\bb\theta)^{\frac{\beta_2}M}=
	&=\lim_{N\to\infty}\lim_{n_1\to-\beta_2}\lim_{n_2\to \frac{\beta_2}M}\frac1{n_1 N}\mE_{\mathcal D} \log \int d\bb W \exp\big(-\frac{\beta_2N\epsilon_W}{2}\sum_{i\mu} W_{i\mu}^2\big)  Z_{\bb W}^{n_1}\prod_{A=1}^M Z_{\bb W}(\bar{\bb\sigma}^A)^{n_2}=\\
 &=\lim_{N\to\infty}\lim_{n_1\to-\beta_2}\lim_{n_2\to \frac{\beta_2}M}\lim_{n_3\to0}\frac1{n_1 n_3 N} \log \mE_{\mathcal D}\Big[\int d\bb W \exp\big(-\frac{\beta_2N\epsilon_W}{2}\sum_{i\mu} W_{i\mu}^2\big)  Z_{\bb W}^{n_1}\prod_{A=1}^M Z_{\bb W}(\bar{\bb\sigma}^A)^{n_2}\Big]^{n_3}.
	\end{split}
\end{equation}
where in the last line we used the replica trick to exchange the order of the logarithm and the expectation over the different realizations of the quenched noise. Now, according to the standard statistical-mechanics pathway, we aim to obtain an explicit expression for $\bar{ \mathcal F}_M$, by which we can obtain an estimate for a set of macroscopic observables which, in turn, allow us to get a picture of the emerging behavior of the system. To this goal, we first 
compute the argument of the logarithm at fixed integers $(n_1,n_2,n_3)$, next we take the thermodynamic limit $N \to \infty$, then we analytically continue the result at real $n_1,n_2,n_3$ such that $n_1/n_2=-M$ and $n_1=-\beta_2$, and send $n_3\to0$. Finally, we take the large dataset limit $M \gg 1$. 
\par\medskip
Before going further, we stress that the structure of the dataset we want to deal with plays a crucial role in the functioning of the BM. In order to make analytical computations within the replica approach, we should specify the reality distribution $Q(\bb\sigma)$. In this paper, we exploit the simplest scenario, where the examples are noisy realizations of a single hidden pattern, and provide them to the network during the training procedure (in a setup close to the product-of-expert approach \cite{hinton2002training}). 
Let us call $\bb\xi$ the pattern we want to model, for which $\bar {\bb \sigma}^A\in \{-1,1\}^N$ with $A=1, ..., M$ are the sample vectors. Specifically, we generate them as $\bar \sigma_i ^A= \chi^A_i \xi_i,$ with $\xi_i=\pm 1$ and $\chi^A_i$ {i.i.d.} random variables distributed as
$$P(\chi^A_i =\pm1)=\frac{1\pm r}{2}.$$ 
From now on, we will refer to $r$ as the {\it quality} of the dataset, as it is the expected fraction of entries of the vector $\bar{\bb\sigma}^A$ correctly aligned to the pattern $\bb \xi$. Within this scenario, the average over all possible realizations of the dataset is simply the expectation value w.r.t. the $\bb \chi$ variables, that is $\mE_{\mathcal D}\to \mE_{\bb \chi}$. Crucial empirical quantities underlying the dataset are, respectively, the $i$-site mean and the empirical overlap between pairs of examples, namely
\begin{eqnarray}\label{eq:moments}
	\sigma_i ^E &=& \frac 1{M} \sum_{A=1}^M \bar\sigma_i ^A,\\
	q^{E}_{AB}&=& \frac1{N_\sigma}\sum_{i=1}^{N_\sigma}\bar \sigma_i ^A \bar \sigma_i ^B.
\end{eqnarray}
Applying the central limit theorem, $\sigma_i ^E  \sim\xi_i r(1+	\sqrt{\rho}z)$ as ${M \gg 1}$, with $z\sim \mathcal N(0,1)$, and
\begin{equation}
\rho := \frac{1-r^2}{r^2M},
\end{equation}
which can be interpreted as the dataset entropy quantifying the amount of information contained in the dataset \cite{alemanno2023supervised}. On the other hand, in the large $N_{\sigma}$ limit, for the empirical overlap one gets
$$
q_{AB}^E =\delta_{AB}+(1-\delta _{AB})r^2,
$$
so that the quality encodes the information about the empirical overlap of the dataset. We anticipate that, in the following calculations, it will be convenient to define $\rho_0  =\rho M=(1-r^2)/r^2\ge0$ and the interesting regime to look at is the one where $\rho_0 \geq 0$ leading to $0<r\le1$.
\par\medskip
 \begin{table}
\begin{center}
\begin{tabular}{llp{2.2cm}p{2.2cm}p{2.2cm}} 
  \toprule
\multirow{2}{*}{}Type&Layer&\multicolumn{3}{c}{Indices}\\
\cmidrule(lr){3-5}
 & & Neurons & Examples & Replicas \\
\midrule
   \multirow{2}{*}{Clamped} &Visible: $\bar \sigma_i^A$ \, \,\, \ & $i=1,\dots,N_\sigma$ & $A=1,\dots,M$ & $a=1,\dots,n_1$ \\
&Hidden: $\bar s_{\mu}^{(A\bar a)\gamma}$ & $ \mu= 1,\dots,N_s$ & & $\bar a=1,\dots,n_2$  \\
%\cline{1-2}
\cmidrule(lr){1-2}
%\midrule
\multirow{2}{*}{Free} &Visible: $\sigma_i^{a\gamma}$ &&   &$\gamma=1,\dots,n_3$   \\
&Hidden: $ s_{\mu}^{a\gamma}$ &  & \\
\bottomrule
\end{tabular}
\end{center}
\caption{{\bfseries Summary of notation used in the replica computations.}
}\label{tab_not}
\end{table}
Before starting the calculations, let us set up the notation (see also Tab.~\ref{tab_not}): visible and hidden units are denoted respectively by $\bb\sigma$ and $\bb s$, with $i=1,\dots,N_\sigma$ and $\mu=1,\dots,N_s$ labelling single units; replica indices for the free units are denoted with lowercase Latin letters, e.g. $a,b=1,\dots,n_1$, and clamped units with barred ones, i.e. $\bar a,\bar b=1,\dots,n_2$; the third replica indices are denoted with $\gamma=1,\dots,n_3$; finally, the examples are labelled with uppercase Latin letters, e.g. $A=1,\dots,M$, and the empirical data are denoted with $\bar{\bb\sigma}^A$. With this notation in mind, we can introduce the order parameters
\begin{eqnarray}\label{eq:OP_visible}
	q_{ab}^\gamma&=&\frac1{N_\sigma} \sum_{i=1}^{N_\sigma} \sigma_i ^{a\gamma} \sigma_i ^{b\gamma},\\ \label{eq:OP2_visible}
	m _{aA}^\gamma &=& \frac1{N_\sigma} \sum_{i=1}^{N_\sigma }\sigma_i ^{a\gamma}\bar \sigma_i^A,
\end{eqnarray}
for the visible layer activity, and
\begin{eqnarray}\label{eq:OP_hidden}
	p_{ab}^\gamma&=& \frac1{N_s} \sum_{\mu=1}^{N_s} s_\mu^{a\gamma} s_\mu ^{b\gamma},\\
 \label{eq:OP2_hidden}	
 n_{a (A\bar a)}^\gamma &=& \frac1{N_s} \sum_{\mu=1}^{N_s} s_\mu ^{a\gamma} \bar s_\mu ^{(A\bar a)\gamma},\\
 \label{eq:OP3_hidden}
	\bar p _{(A\bar a)(B \bar b)}^	\gamma&=& \frac1{N_s} \sum_{\mu=1}^{N_s}\bar s_\mu^{(A\bar a)\gamma} \bar s_\mu ^{(B \bar b)\gamma},
\end{eqnarray}
for the hidden one. Concerning the visible layer, the first order parameter \eqref{eq:OP_visible} is the usual overlap order parameter for spin-glass systems, measuring the level of self-organizing behavior of independent replicas, and it is a $n_1\times n_1$ matrix for each $\gamma$. The second one \eqref{eq:OP2_visible} is the Mattis magnetization of with examples, and measures the overlap of the visible units' configuration with the dataset. Thus, it measures the capabilities of the BM to generate samples according to the same statistics of the dataset $\mathcal D$. They are $n_3$ matrices of dimensions $M\times n_1$. The $ p^\gamma_{ab}$, $n_{a(A\bar a)}^\gamma$ and $\bar p_{(A\bar a)(B\bar b)}^\gamma$ overlaps are resp. $n_1\times n_1$, $n_1\times Mn_2$ and $M n_2\times Mn_2$ matrices for each $\gamma$. The first \eqref{eq:OP_hidden} and the third \eqref{eq:OP3_hidden} order parameters measure the overlap between independent replicas resp. if the BM is let free to evolve towards equilibrium or rather clamped to the data. Conversely, the second order parameter \eqref{eq:OP2_hidden}  %$n_{a(A\bar a)}^\gamma$ instead 
measures the correlation between independent replicas of the hidden layer within the free and the clamped case. After some lengthy computations, one finds that the argument of the logarithm in \eqref{eq:tot_fe} reads as 
\begin{equation}\label{eq:general_replica_Z}
	\begin{split}
		&\mE _{\bb\chi}	\Big[\int d\bb W \exp\big(-\frac{\beta_2N\epsilon_W}{2}\sum_{i\mu} W_{i\mu}^2\big) Z_{\bb W}^{n_1}	\prod_{A=1}^M Z_{\bb W}(\bar{\bb\sigma}^A )^{n_2}\Big]^{n_3}=\\&
		=\int  \frac{d \bb O}{\mathcal V}\exp\Big[-\frac{\eta (1-\eta)N}{2\beta_2 \epsilon_W}\sum_{\gamma ab} q^\gamma_{ab} p^{\gamma}_{ab}-\frac{\eta(1-\eta)N}{\beta_2 \epsilon_W}\sum_{\gamma a A \bar a} m^\gamma_{aA} n^{\gamma}_{a(A\bar a)}-\frac{(1-\eta)N}2 \sum_\gamma \log \text{det} \big(\bb 1- \frac{\eta {\bb q}^{\gamma}}{\beta_2 \epsilon_W } \big)\\&-\frac{(1-\eta )N}2  \sum_\gamma \log \text{det} \Big(\bb 1- \frac{2\, {\bb r}^{\gamma}}{1-\eta } \Big)\Big]\mE_{\bb \chi}\sum_{\bb\sigma} \exp\Big(
		\frac{1-\eta}{\beta_2 \epsilon_W} \sum_{\gamma a A \bar ai} n^\gamma_{a(A\bar a)}\sigma_i ^{a\gamma}\bar \sigma_i ^A+\frac{1-\eta}{2\beta_2 \epsilon_W} \sum_{\gamma ab i}p^{\gamma }_{ab}\sigma_i ^{a\gamma}\sigma_i^{b\gamma}
		\Big),
	\end{split}
\end{equation}
where $\eta=N_\sigma/N$, and we set $\beta_1=1$ (without loss of generality, as it can be trivially restored a posteriori by suitably rescaling the other parameters), $\mathcal V$ is a volume factor in which we absorbed all the unessential constant terms,
$\boldsymbol{r}$ is the effective overlap between hidden clamped replicas
\begin{equation}\label{eq:r_true}
		r^\gamma_{(A\bar a)(B\bar b)}:=\frac{\eta(1-\eta)}{2\beta_2\epsilon_W}	q_{AB}^E+\frac{\eta^2(1-\eta)}{2\beta_2 \epsilon_W}\sum_{ab} m ^\gamma_{aA} \big(\bb 1 - \frac{\eta}{\beta_2\epsilon_W}{\bb q}^\gamma\big)^{-1}_{ab} m^\gamma_{bB},
\end{equation}
and finally $\boldsymbol{O}$ is the set of all order parameters, i.e.
\begin{equation}
     O_{(A\bar a)(B\bar b)}=\{q_{ab}^\gamma,q_{ab}^\gamma,m_{aA}^{\gamma},n_{a(A\bar a)}^{\gamma},\bar p_{(A\bar a)(B\bar b)}^\gamma\}.
\end{equation}
The details of the replica computations yielding Eq. \eqref{eq:general_replica_Z} are reported in App. \ref{app:details_replica}. To proceed with the evaluation of \eqref{eq:general_replica_Z}
 we need to make an assumption on the structure of the order parameters, as detailed in the next subsection.
 
 \subsection{Replica-symmetric Ansatz and self-consistency equations}
The scenario we address in the paper corresponds to the RS assumption for the order parameters. For the overlaps, this leads to the prescription
\begin{eqnarray}
	q_{ab}^\gamma&=&\delta_{ab} +(1-\delta_{ab})q,\label{eq:RS_overlaps1}\\
	p_{ab }^\gamma&=& \delta_{ab}P+(1-\delta_{ab})p\label{eq:RS_overlaps2}.
\end{eqnarray}
As for the magnetizations, we assume that their value is replica-symmetric {\it and} independent of the sample index, meaning that all examples in the dataset are assumed to display the same overlap with the final configuration of the visible layer. This means that
\begin{eqnarray}
	m^\gamma_{aA} &=& m,\label{eq:RS_magn1}\\
	n^\gamma_{a(A\bar a)}&=&n\label{eq:RS_magn2}.
\end{eqnarray}
A more general form of the RS snsatz in which the single example index $A$ is preserved is discussed in App. \ref{app:general_RS} and shown to be recastable (in the interesting regimes) in the current one.\par\medskip
By plugging \eqref{eq:RS_overlaps1}-\eqref{eq:RS_magn2} into \eqref{eq:tot_fe} we get
\begin{equation}
	\begin{split}
\bar 	{	\mathcal F}%&=\lim_{N\to\infty}\lim_{n_1\to-\beta_2}\lim_{n_2\to \frac{\beta_2}{M}}\lim_{n_3\to 0}\frac1{n_3n_1N}\log \mE _{\bb\chi}	\Big(\mE_{\bb W}Z(\bb W)^{n_1}	\prod_{A=1}^M Z(\bar{\bb\sigma}^A ,\bb W)^{n_2}\Big)^{n_3} \\
&=\lim_{M\to\infty}\lim_{N\to\infty} \lim_{n_1\to-\beta_2}\lim_{n_2\to \frac{\beta_2}{M}}\lim_{n_3\to 0}\frac1{n_3n_1N} \log \frac1{\mathcal V} \int d \bb O \exp\big(Nn_3 f_M(\bb O)\big),
	\end{split}
\end{equation}
with
\begin{equation}\label{eq:fe_exp}
	\begin{split}
	\frac{f_M(\bb O)}\eta
	&= - \frac{\alpha\beta_\epsilon}{2\beta _2 } n_1(n_1-1)pq- \frac{\alpha\beta_\epsilon}{\beta _2 } n_1 n_2 M m n -n_1\frac{\alpha\beta_\epsilon}{2\beta_2} p
	-\frac{\alpha}2(n_1-1)\log\Big[1-\frac{\beta_\epsilon}{\beta_2}(1-q)\Big]+\\
	&
	-\frac{\alpha}{2}\log \Big\{\Big[1-\frac{\beta_\epsilon}{\beta_2 }{Mn_2}r^2{(1+\rho)}\Big]\Big[{1-\frac{\beta_\epsilon}{\beta_2 }(1-q)-n_1 \frac{\beta_\epsilon}{\beta_2 }q}\Big]-\frac{\beta_\epsilon^2 }{\beta_2^2}{Mn_2n_1 m^2}\Big\}+
	\\&
	+  \mE_z\log \mE_\tau\cosh^{n_1}\Big[Mn_2 \frac{\alpha\beta_\epsilon}{\beta_2}n r ({1+\sqrt \rho z})+\sqrt{\frac{\alpha\beta_\epsilon}{\beta_2}p}\tau\Big],
	\end{split}
\end{equation}
where we also defined $\alpha = N_s /N_\sigma=( {1-\eta})/\eta$, $\epsilon_W= \eta \beta_\epsilon^{-1}$ and $r=(1+\rho _0)^{-1/2}$. 
\par\medskip
At this point, we perform the analytical continuation to real number of replicas, and take the limits $n_1\to-\beta_2,n_2 \to \beta_2/M,n_3\to0$, next, we send $M\to\infty$ with fixed $\rho_0\ge 0$. The extremization of the function \eqref{eq:fe_exp} %after taking the limits $n_1\to-\beta_2,n_2 \to \beta_2/M$ 
leads to the self-consistency equations
\begin{eqnarray}
		\tilde n &=& \frac{T_2}{m^2 -\Big(\frac{1}{1+\rho_0}-\frac1{\beta_\epsilon}\Big)\big[q-T_2(1-q)+\frac1{\beta_\epsilon}\big]},\label{eq:n_final_0}\\
	\tilde p &=& \frac{T^2_2}{\frac1{\beta_\epsilon}-T_2 (1-q)}+\Big(\frac{1}{1+\rho_0}-\frac1{\beta_\epsilon}\Big) \tilde n T_2,\label{eq:p_final_0}\\
		m &=& \frac{\tilde m}{\sqrt{1+\rho_0}},\label{eq:m_final_0}\\
		q &=& I_2,\label{eq:q_final_0}
\end{eqnarray}
where $T_2=\beta_2^{-1}$, $\tilde m= I_1$ and we rescaled for convenience $n = \tilde n m /\beta_\epsilon$ and $p=\tilde p/(\beta_\epsilon T_2)$, with the integrals $I_1$ and $I_2$ defined as
\begin{eqnarray}
	I_1 &=& \frac{\int d\mu(\tau) \cosh^{-\beta_2}g(\tau)\tanh g(\tau)}{\int d\mu(\tau)\cosh^{-\beta_2}g(\tau)},\nonumber\\
	I_2 &=& \frac{\int d\mu(\tau)\cosh^{-\beta_2}g(\tau)\tanh^2 g(\tau)}{\int d\mu(\tau)\cosh^{-\beta_2}g(\tau)},\nonumber
\end{eqnarray}
with $d\mu(\tau)=(2\pi)^{-1/2}\exp(-\tau^2/2)d\tau$ being the Gaussian measure, and 
\begin{equation}\label{eq:g-function}
g(\tau)= \frac{ \alpha \tilde n m}{\sqrt{1+\rho_0}}+\sqrt{\alpha \tilde p}\tau.
\end{equation}
We further stress that the quantity $\tilde m$ is precisely the (thermodynamic value of the) magnetization of the visible layer activity with respect to the pattern $\bb\xi$. This can be understood by adding to the energy function \eqref{eq:H} a source term of the form $\theta \sum_i \xi_i \sigma_i$, with $\theta$ sent to $0$ at the end of the computation. The effect of this source term is simply to add a factor $\theta$ in the $\cosh$ function appearing in the free energy, then defining $\tilde m = \partial \bar {\mathcal F}/\partial \theta\vert_{\theta=0}$ (as standard in statistical mechanics) we precisely identify this magnetization $\tilde m=I_1$ with the ground-truth magnetization. Eqs. \eqref{eq:n_final_0}-\eqref{eq:q_final_0} can be numerically solved for any choice of the tunable parameters $\alpha,\beta_\epsilon,T_2,\rho_0$. However, before discussing the solutions of the self-consistency equations, we move our attention to a peculiar point of the RS ansatz, see Eqs.~\eqref{eq:RS_overlaps1}-\eqref{eq:RS_magn2}.

\subsection{Extremization conditions}\label{subsec:theo_cons} 
When performing the analytical continuation to real values of $(n_1,n_2,n_3)$, the logarithmic contributions to the free energy acquire crucial roles. In particular, since the relevant limit is taken under the condition $n_2M=-n_1=\beta_2$, one can promote these two replica numbers to complex values as $n_1=\beta_2 e^{i\phi}$ and $n_2=\beta_2 e^{i(\phi-\pi)}$, and then sending $\phi \to\pi^-$. In doing this, one of the two variables would hit the branch cut of the logarithm (which can be chosen to lie on the semi-axis of negative reals). This means that the resulting free energy would acquire a non-vanishing imaginary part, unless the following two conditions are satisfied (which equals asking that the imaginary contributions are exactly zero):
\begin{eqnarray}
\textrm{I)}  \quad  & & \quad 1-\frac{\beta_\epsilon}{\beta_2}(1-q)>0, \label{eq:RS_cond1}\\
\textrm{II)} \quad  & & \quad   \Big(1- \frac{\beta_\epsilon}{1+\rho _0}\Big)\Big(1-\frac{\beta_\epsilon}{\beta_2}q+\beta_\epsilon q\Big)+\beta_\epsilon^2 m^2>0. \label{eq:RS_cond2}
\end{eqnarray}
If these conditions are simultaneously satisfied, the quenched free energy is real, and the RS solution is a proper extremal point for the free energy. In the following, we examine the consequences of these conditions for the three relevant solutions of the saddle point equations, namely the paramagnetic, spin-glass and retrieval solutions.
\begin{itemize}
    \item \emph{Ergodic solution ($m=q=0$). } Conditions I and II in this setting reduce to
$$
T_2 <\beta_\epsilon^{-1} \quad \text{and}\quad \beta_\epsilon<1+\rho_0.
$$
Hence, for $\beta_\epsilon<1+\rho_0$ (from now on this regime will be called the \textit{high-regularization regime}\footnote{In fact, we recall that $\beta_{\epsilon}$ is related to the regularization parameter $\epsilon_W$ introduced in \eqref{eq:H_W} by $\epsilon_W=\eta \beta_{\epsilon}^{-1}$.} and the line $\beta_\epsilon=1+\rho_0$ will be called \textit{regularization threshold}) the paramagnetic solution exists only at low training temperature (i.e., $T_2 <1/\beta_\epsilon$). In the opposite case $\beta_\epsilon>1+\rho_0$, the paramagnetic solution ceases to exist at any value of the temperature $T_2$.\footnote{We recall that the temperature $T_2$ controls the thermal level in the Langevin dynamics of the training procedure, while the Boltzmann-Gibbs distribution of the visible layer is tuned by the temperature $T_1$ (which we set to 1 in our replica computations). Resuming the sampling temperature $T_1$ with $\beta_\epsilon\to \beta_1^2 \beta_\epsilon$, it is clear that the paramagnetic solution does exist for $
T_1 >\text{max}\Big(\sqrt{\beta_\epsilon T_2}, \sqrt{\frac{\beta_\epsilon}{1+\rho_0}}\Big).$}
\item \emph{Spin-glass solution ($m=0$ and $q>0$).} 
Let us first consider the $\beta_\epsilon <1+\rho_0$ case, where conditions I and II reduce to
$$
\text{I}:\begin{cases}
	q	>0 \quad \quad\quad \quad\text{if } T_2 <\beta_\epsilon^{-1}\\
	q	>\frac{\beta_\epsilon-\beta_2}{\beta_\epsilon}\quad\quad \text{if } T_2 >\beta_\epsilon^{-1}
\end{cases},
 \quad \quad \quad
\text{II}:\begin{cases}
	q	>0 \quad \quad\quad \quad\text{if } T_2 <\beta_\epsilon^{-1}\\
	q	>\frac{\beta_\epsilon-\beta_2}{\beta_\epsilon(\beta_2+1)}\quad \text{if } T_2 >\beta_\epsilon^{-1}
\end{cases}.
$$
This means that, if $T_2<\beta_\epsilon^{-1}$, the conditions I and II are trivially satisfied for the spin-glass solution. On the other hand, if $T_2>\beta_\epsilon^{-1}$, this solution can still exist, provided that $q>(\beta_\epsilon-\beta_2)/\beta_\epsilon>0$, thus ruling out the existence of critical transitions, as indeed found in App. \ref{app:critical}. Conversely, if $\beta_\epsilon>1+\rho_0$, we have
$$
\text{II}:\begin{cases}
	\text{never} \quad \quad\quad \ \ \text{if } T_2 <\beta_\epsilon^{-1}\\
	q	<\frac{\beta_\epsilon-\beta_2}{\beta_\epsilon(\beta_2+1)}\quad \text{if } T_2 >\beta_\epsilon^{-1}
\end{cases}.
$$
The condition I does not depend on $\rho_0$, thus it is the same as the previous case. One can see that it is impossible to satisfy conditions I and II simultaneously in the low regularization regime (i.e. where $\beta_{\epsilon}>1+\rho_0$), so even the spin-glass solution ceases to exist there. 
\begin{figure}[h!]
	\centering
	 \includegraphics[width=0.6\textwidth]{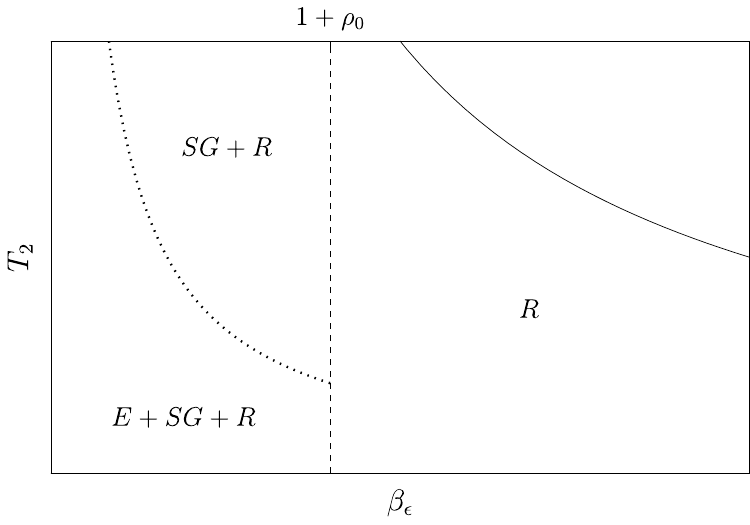}
	\caption{{\bfseries Existence regions for the solutions extremizing the free energy.} The figure illustrates the existence conditions for the ergodic (E), spin-glass (SG) and retrieval (R) replica-symmetric solutions in the $(\beta_\epsilon,T_2)$ plane. The vertical dashed line corresponds to $\beta_\epsilon=1+\rho_0$, identifying the regularization threshold (above which neither the ergodic nor the spin-glass solutions exist), while the dotted one is $T_2 = \beta_\epsilon^{-1}$ (separating the region in which ergodic solution is a proper extremal point for the free energy). The solid line corresponds to $T_2 = (1+\rho_0)/(\beta_\epsilon \rho_0)$: below this curve (and above the regularization threshold), only the retrieval RS solution exists. Above the solid line, no RS solution is found, as we deepen in the next sections.
 }\label{fig:regions}
\end{figure}
\item \emph{Retrieval solution ($m, q>0$).} Unlike the ergodic and the spin-glass solutions, for which we proved the existence of regions where they are not proper extremal points for the free-energy, we now show that the retrieval solution can exist above and below the regularization threshold.
If $\beta_\epsilon<1+\rho_0$, the condition II holds for all $m, q>0$ (provided that condition I is satisfied) as it is the sum of positive quantities. In the opposite case $\beta_\epsilon>1+\rho_0$, condition II becomes
$$
m^2 > \frac1{\beta_\epsilon^2}\Big(\frac{\beta_\epsilon}{1+\rho_0}-1\Big)\Big[1-\frac{\beta_\epsilon}{\beta_2}(1-q)+\beta_\epsilon q\Big]>0,
$$
again under the hypothesis that condition I is satisfied. In particular, if $T_2 >\beta_\epsilon^{-1}$, it again implies
$$
q > \frac{\beta_\epsilon -\beta_2}{\beta_\epsilon},
$$
meaning that $m$ and $q$ must be strictly positive in this regime, again forbidding the presence of critical transitions. This is in agreement with our analysis in App. \ref{app:critical}. Finally, we notice that, in the retrieval region, another peculiar relation between the control parameters holds. Indeed, when the system works properly as a generative model, the overlap of the visible units displays the natural bound $q\le r^2=(1+\rho_0)^{-1}$ (meaning that the network is properly sampling according to the dataset statistics, thus capturing the empirical overlap $r^2$ between generated examples), which directly implies 
$$1-\frac{\beta_\epsilon}{\beta_2}(1-q)\le 1-\frac{\beta_\epsilon}{\beta_2}\big(1-\frac1{1+\rho_0}\big). $$ 
Thus, in order for condition I to be fulfilled, it is necessary that the r.h.s. of this inequality is strictly positive, which leads to
\begin{equation}
    \label{eq:rs_break}
    T_2<\frac{1+\rho_0}{\beta_\epsilon \rho_0}.
\end{equation}
\end{itemize}
A summary of these results is depicted in Fig. \ref{fig:regions}. 
The previous discussion highlights the working regimes of the model, in particular, we were able to find relations between the parameters $T_2$, $\beta_\epsilon$ and $\rho_0$ differentiating relevant regimes in the generative capabilities of the model. The relation $T_2=\beta_\epsilon^{-1}$ in fact separates low and high training temperature phases. The other crucial combination of control parameters is the regularization threshold $\beta_\epsilon=1+\rho_0$: above this line, spin-glass and paramagnetic solution are not present, and only the retrieval one survives (at least in the current RS description). Further, since above the regularization threshold the only RS solution is the ansatz \eqref{eq:RS_overlaps1}-\eqref{eq:RS_magn2} (as deepened in App. \ref{app:general_RS}), we are left to understand the role of the curve $T_2 = (1+\rho_0)/(\beta_\epsilon\rho_0)$, which we will deepen in the following sections. 

\section{Solution of the self-consistency equations} \label{sec:selfcons}
We now move to present the numerical solutions of the self-consistency equations under the RS ansatz \eqref{eq:RS_overlaps1}-\eqref{eq:RS_magn2}, exploring the parameter space $(\beta_\epsilon,T_2,\rho_0)$. We start with the low-regularization regime and explore the nature of the solution below the regularization threshold, showing that also paramagnetic and spin-glass solutions are well-defined here, with the retrieval one being only a local minimum for the quenched free energy. We also show that critical transitions occur in this regime and finally, we compare the phenomenology of the low and high regularization regimes.

\subsection{Low-regularization regime: $\beta_\epsilon >1+\rho_0$}
We start by investigating the retrieval solution in the low-regularization regime, {i.e.} low $\epsilon_W$ or equivalently high $\beta_\epsilon$.  

\begin{figure}[t!]
	\centering
	\includegraphics[width=\textwidth]{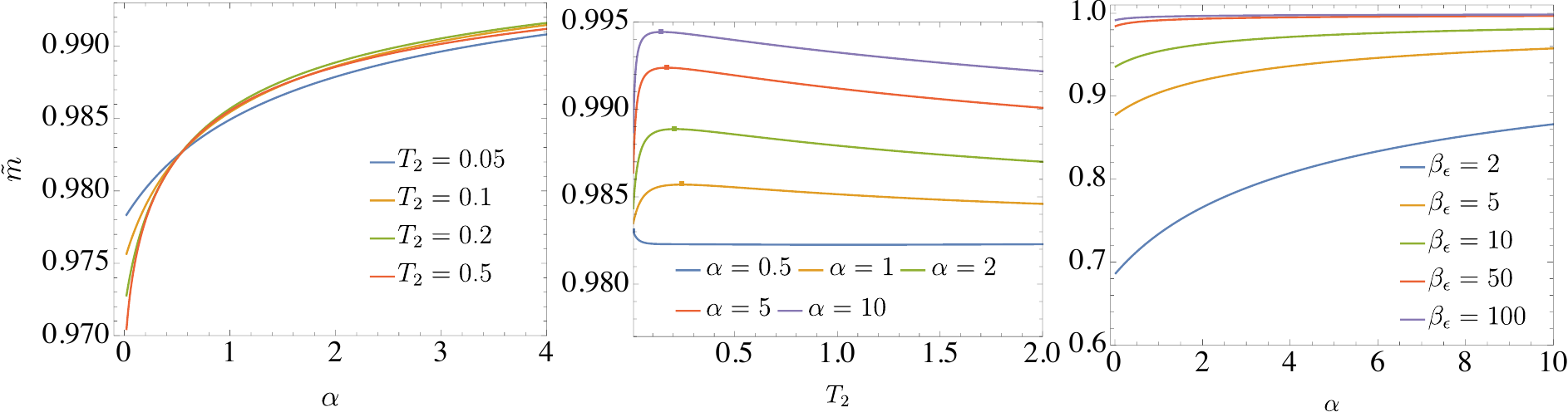}
	\caption{{\bfseries Numerical solutions of the self-consistency equations for $\rho_0=0$}. The three plots show details of the numerical solutions for the order parameter $\tilde m$ of the self-consistency equations \eqref{eq:n_final_0}-\eqref{eq:q_final_0}. In the left plot, we reported the results for $\tilde m$ as a function $\alpha$ for various values of $T_2$ at $\beta_\epsilon=30$. In the plot in the center, we plotted the magnetization $\tilde m$ as a function of $T_2$ for various values of $\alpha$ and $\beta_\epsilon=30$, highlighting the best generating performances with square markers. In the right plot, we finally reported the magnetization at $T_2=0.1$ as a function of $\alpha$ for various values of the regularization parameter $\beta_\epsilon$.}\label{fig:rho0_solutions}
\end{figure}

\paragraph{The zero-entropy limit: $\rho_0=0$.}
This limit can be reached, regardless of $M$, by taking $r = 1$, in such a way that the collection of $M$ examples is fully redundant: this special case can therefore be considered as a test to highlight the intrinsic working regimes of the model. A further property of this limit is that, according to Eq. \eqref{eq:m_final_0}, the magnetization of the sampled configuration with the examples in the dataset is the same as the one with the true feature, i.e. $m=\tilde{m}$. The behavior of the solution of the self-consistency equations as the external parameters $\alpha$, $T_2$ and $\beta_\epsilon$ are varied is reported in Fig. \ref{fig:rho0_solutions}. 
Given the simplicity of the task in this regime -- sampling according to the dataset statistics, characterized by a unique ground-truth direction $\bb\xi$ -- the system is always able to retrieve information, with performances depending on the values of the external parameters. In the plot on the left, we report the magnetization as a function of $\alpha$ by varying $T_2$ at $\beta_\epsilon=30$, and we see that the system always exhibits high sampling capabilities ($m \gtrsim0.97$). Also, as expected, the more parametrized the network (i.e., increasing $\alpha$), the better the sampling w.r.t. the desired statistics. However, by inspecting the plot more carefully, it can be noticed that the behavior in $T_2$ is not trivial, as the best magnetization for fixed $\alpha$ does depend on $T_2$. This behavior is indeed clear by looking at the middle plot, in which we compute the final magnetization as a function of $T_2$ for various values of $\alpha$. In particular, at high $\alpha$, the best sampling performance is achieved at $T_2\approx0.2$, while for relatively low values of $\alpha$, the best magnetization is achieved with a deterministic training (i.e. $T_2=0$). Remarkably, since stochastic noise is usually realized by averaging the gradients over small portions of the available dataset (i.e. the stochastic gradient descent approach), $T_2$ is strictly related to the size of the mini-batches. Thus, the results in the middle plot provide a prescription for a optimal choice of the mini-batch size given the other parameters.
Finally, the right plot shows the behavior of the final magnetization as a function of $\alpha$ at fixed $T_2 =0.1$ for various values of $\beta_\epsilon$. These results state an expected behavior: for low $\beta_\epsilon$, the regularization is strong, meaning that the quadratic potential (namely, the $L_2$-regularization) in the network configuration space is narrow; this implies that, during training, the system is strongly constrained to move around the trivial configuration $\bb W=0$, thus it has not enough freedom to explore the free energy landscape and reach equilibrium states with high values of the magnetization. Increasing $\beta_\epsilon$ would result in a gradual reduction of the importance of the regularization, thus the system better explores the landscape to reach good equilibrium configurations in terms of sampling performances.

\paragraph{Finite $\rho_0$ regime.}
\begin{figure}[t!]
	\centering
	\includegraphics[width=\textwidth]{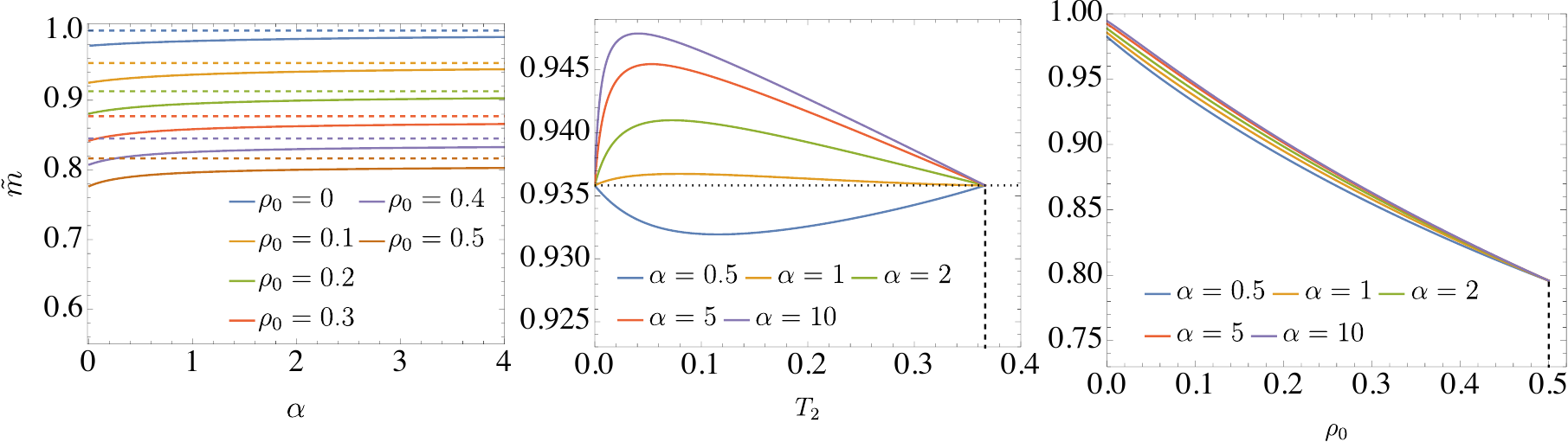}
	\caption{{\bfseries Numerical solutions of the self-consistency equations for $\rho_0\neq0$}. The three plots show details about the numerical solutions for the order parameter $\tilde m$ of the self-consistency equations \eqref{eq:n_final_0}-\eqref{eq:q_final_0}. In left plot, we reported the results for $\tilde m$ as a function $\alpha$ at $T_2 =0.05$ for various values of $\rho_0$ at $\beta_\epsilon=30$. The dashed lines corresponds to the upper bound given by $\tilde m=r$. In the plot in the center, we plotted the magnetization $\tilde m$ as a function of $T_2$ for various values of $\alpha$ at $\rho_0=0.1$ and $\beta_\epsilon=30$. In the right plot, we finally reported the magnetizations a function of $\rho_0$ for various values of the parametrization load $\alpha$ at $T_2=0.1$ and $\beta_\epsilon=30$.}\label{fig:mvsrho0}
\end{figure}

\begin{figure}[t!]
	\begin{minipage}{\textwidth}
		\centering
		\includegraphics[width=\textwidth]{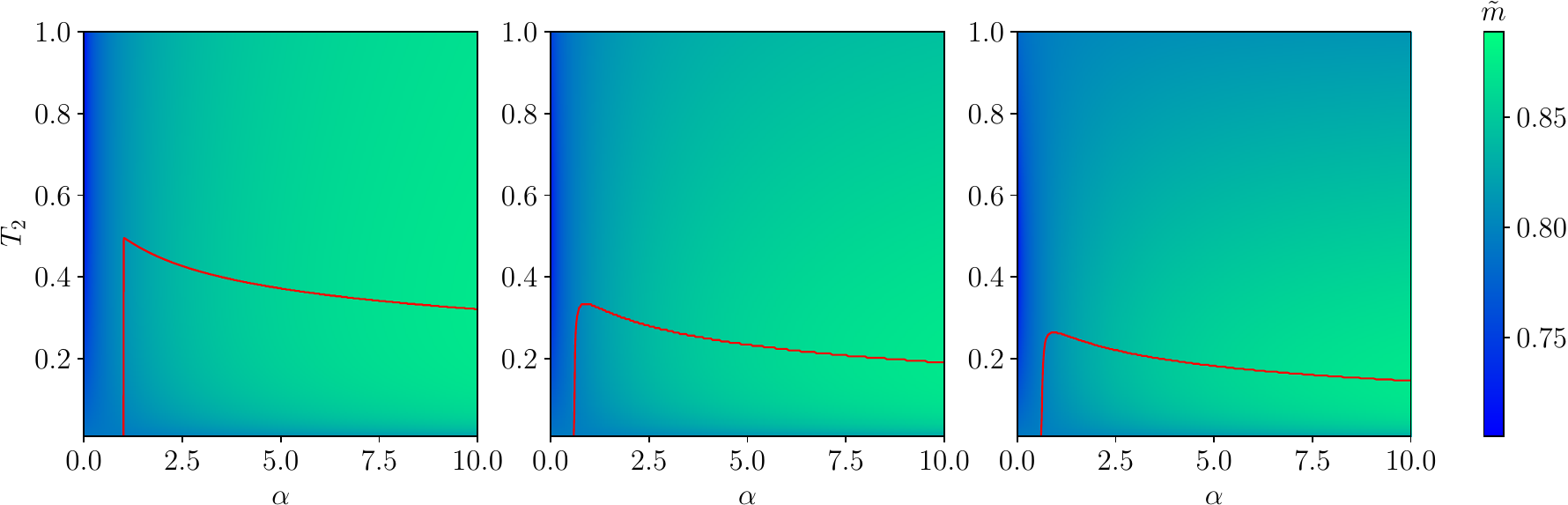}
	\includegraphics[width=\textwidth]{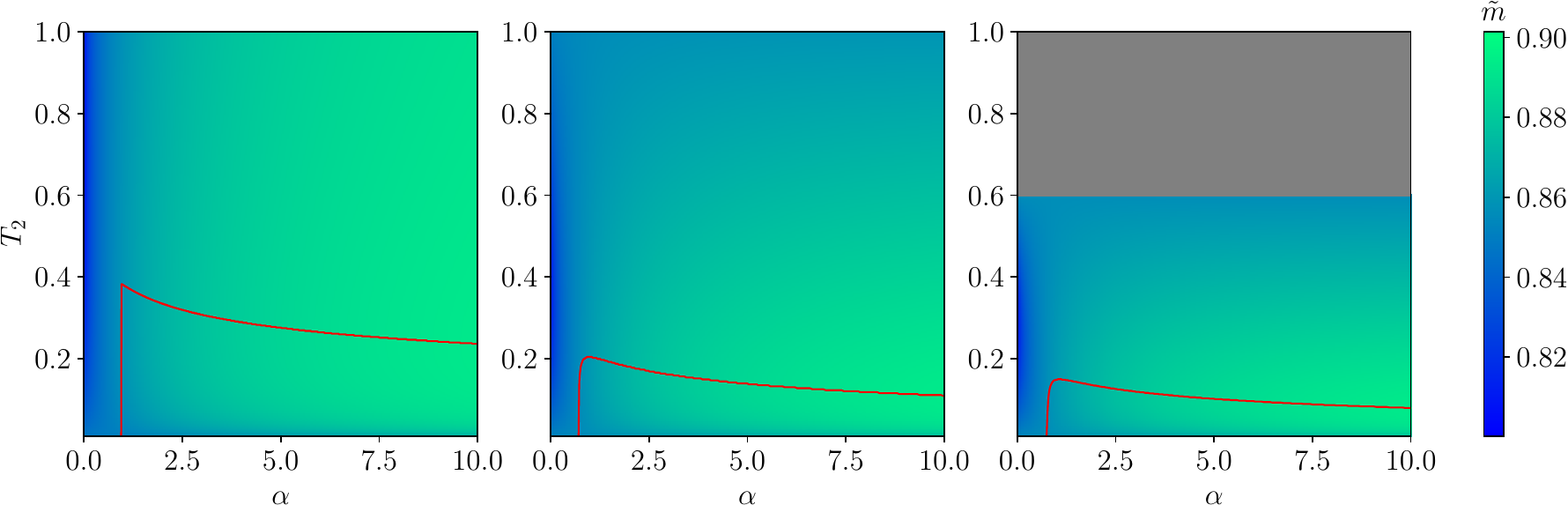}
	\includegraphics[width=\textwidth]{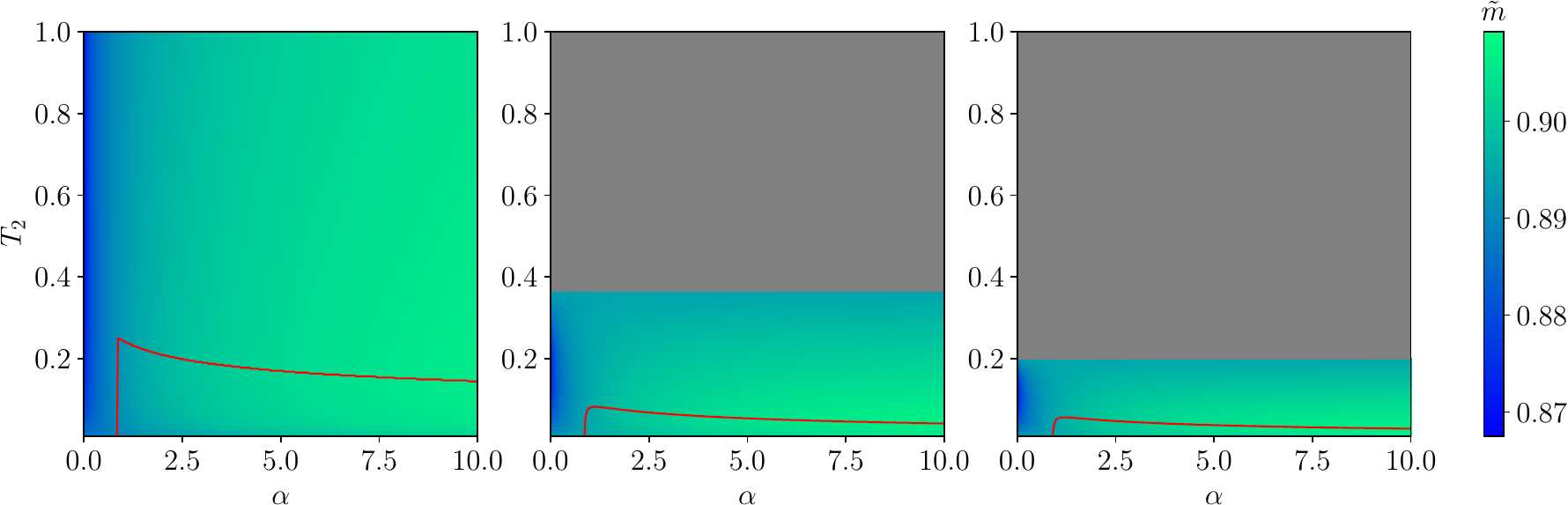}
\end{minipage}
	\caption{{\bfseries Response diagram of the RBM.} The panels show the response of the RBM in terms of the final overlap $\tilde m$ of the visible layer with the pattern $\bb\xi$ in the $(\alpha,T_2)$ plane for various values of $\beta_\epsilon$ and $\rho_0$. The colour map illustrates the values taken by the magnetisation $\tilde m$ (the legend is reported on the right) and the red curve coincides with the 1-dimensional subspace at which the maximum magnetization $\tilde m$ is achieved. The panels are organized as follows: from the left to the right, we increase the dataset entropy $\rho_0$ ($0,0.1,0.2$), while from the top to the bottom we increase the regularization parameter $\beta_\epsilon$ ($=5,10,30$). 
 We stress that Eq. \eqref{eq:rs_break} provides the temperature of RS retrieval solution breaking at $T_2\approx 0.37$ for $\rho=0.1$ and $\beta_\epsilon=30$, $T_2= 0.6$ for $\rho=0.2$ and $\beta_\epsilon=10$, $T_2= 0.2$ for $\rho=0.2$ and $\beta_\epsilon=30$, in perfect agreement with the numerical solution of the self-consistency equations resumed in the color maps.}
	\label{fig:phase2}
\end{figure}
In this case, the value $r=(1+\rho_0)^{-1/2}$ is an upper bound for the final magnetization $\tilde m$, since we are extracting and generating samples according to the (training) dataset statistics: the system would be able to sample configurations which are related to the pattern $\bb \xi$ at most with a correlation $r$. By inspecting the left plot in Fig. \ref{fig:mvsrho0}, this is indeed the case. In this plot, we report the result of the self-consistency equations for $\tilde m$ as a function of $\alpha$ at $T_2=0.05$ for various values of the dataset entropy $\rho_0$. Again, the system is always able to generate samples with non-trivial correlation with the pattern $\boldsymbol{\xi}$, but clearly this overlap will be bounded by $r$ (the dashed lines). The higher $\alpha$, the closer the magnetization $\tilde m$ moves towards this theoretical upper bound. What is interesting in this regime is the behavior of $\tilde m$ as a function of $T_2$. The results are similar to the $\rho_0=0$ case: in particular, for a low parametrization $\alpha$, the function $\tilde m(T_2)$ is convex, while at high $\alpha$ it is concave, and there exists a non-vanishing $T_2$ at which $\tilde m$ is maximum. Interestingly, at $\rho_0>0$ and increasing $T_2$ the magnetization curves collapse to a unique point, above which we do find further RS solutions; more remarkably, the value of the temperature where this coalescence takes place is precisely the r.h.s. of \eqref{eq:rs_break}, thus suggesting that the curve $T_2=(1+\rho_0)/(\beta_\epsilon \rho_0)$ is an intrinsic feature of the RS solution. By inspecting at the plot in the center of Fig. \ref{fig:mvsrho0} (where $\rho_0=0.1$), we see that all of the curves $\tilde m (T_2)$ for different parametrization loads $\alpha=N_s/N_{\sigma}$ have a common origin (at $T_2=0$, $\tilde m \approx 0.935$ regardless of $\alpha$). As $T_2$ is increased, the curves split in a $\alpha$-dependent way: for low parametrization, the magnetization decreases, while for $\alpha$ high enough, it starts to increase. These curves resp. develop a minimum/maximum, then they start to approach themselves again. A similar behavior is also present in the plot for $\tilde m$ versus $\rho_0$. By inspecting at the right plot in Fig. \ref{fig:mvsrho0} at $T_2=0.1$, we see that the order parameter $\tilde m$ is a decreasing function of $\rho_0$, until the curves merge: above this threshold, there is no retrieval solution within the RS approximation, and the value of $\rho_0$ at which the solution breaks down is again consistent with \eqref{eq:rs_break}. Also this behavior can be explained by looking at the self-consistency equations \eqref{eq:n_final_0}-\eqref{eq:q_final_0}: a non-vanishing $\rho_0$ is responsible for a lower signal in the $g$-function \eqref{eq:g-function}, so we expect that -- for high enough $\rho_0$ -- the generative capabilities break down, and the system experiences an abrupt change in the retrieval phenomenology.

\paragraph{A summary of retrieval solution in the low regularized regime.} We resume our results in the $3 \times 3$ grid of response diagrams in the $(\alpha,T_2)$ plane for different values of $\beta_\epsilon$ and $\rho_0$, see Fig. \ref{fig:phase2}. Going from the left to the right, we increase $\rho_0$ ($=0,0.1,0.2$), from the top to the bottom we increase  $\beta_\epsilon$ ($=5,10,30$). The colour map is related to the values of the magnetisation $\tilde m$ (the legend is reported on the right) and the red curve coincides with the 1-dimensional subspace at which the maximum magnetization $\tilde m$ is achieved. {For $\rho_0 >0$, in the gray region (corresponding to the domain in the $(\alpha,T_2)$ space where \eqref{eq:rs_break} is violated) we did not find any RS-retrieval solution (even initializing the fixed point iteration algorithm in more {\it exotic} initial condition, e.g. $q>r^2$). This strongly suggests that the system there experiences RSB effects, as no RS solutions can be found here which properly extremizes the free energy (spin glass and ergodic ones are not present according to our previous discussion). Indeed, conditions \eqref{eq:RS_cond1}-\eqref{eq:RS_cond2} are clearly a direct consequence of the RS structure of the overlap order parameters, resulting in two conditions for the two order parameters $\tilde m$ and $q$. Adopting the Parisi RSB scheme, we would instead have more independent overlap order parameters, resulting in a higher freedom to fulfill proper extremization conditions for the free energy. Then, we argue that, as RS solution fails to satisfy \eqref{eq:RS_cond1}-\eqref{eq:RS_cond2}, the system unavoidably enters a phase in which the only possible minima exhibit RSB. We will deepen this point with numerical simulations.
}

\subsection{Low \emph{versus} high regularization regimes}
\begin{figure}
    \centering
    \begin{minipage}{0.45\textwidth}
    \includegraphics[width=\textwidth]{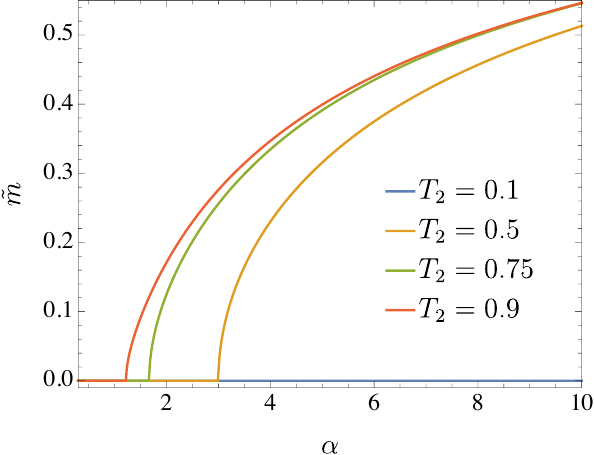}
    \end{minipage}
    \begin{minipage}{0.45\textwidth}
      \includegraphics[width=\textwidth]{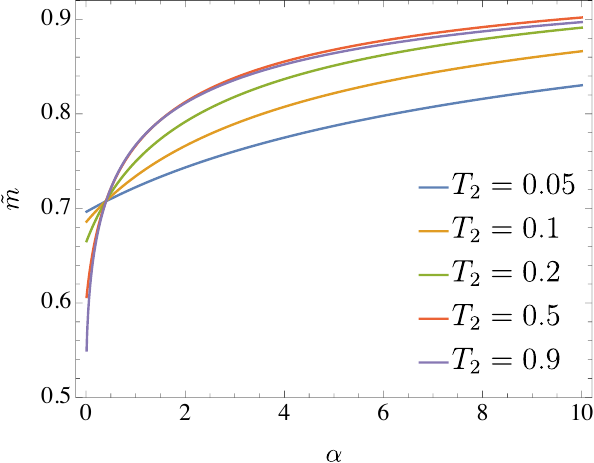}
    \end{minipage}
    \caption{{\bfseries Comparison between retrieval solutions below and above the regularization threshold.} The two plots show the behavior of the retrieval solution (in terms of the feature magnetization $\tilde m$ as a function of the relative width of the hidden layer $\alpha$) below ($\beta_\epsilon =0.5$) and above ($\beta _\epsilon =2$) the regularization threshold at $\rho_0 = 0$ and for various values of the temperature $T_2 <1$.}
    \label{fig:lowVShigh}
\end{figure}

To conclude this Section, we compare the retrieval solutions above and below the regularization threshold (resp. the low and high regularized regimes, as we recall that the $L_2$-weight is proportional to $\beta_\epsilon ^{-1}$). Without loss of generality, we set $\rho_0=0$, so that the two regimes are characterized by $\beta_\epsilon >1$ and $\beta_\epsilon <1$ respectively. In particular, we set $\beta_\epsilon =0.5$ and $\beta_\epsilon = 2$, and solve the self-consistency equations for various values of $T_2$, and compare the curves $\tilde m (\alpha)$. The results are reported in Fig. \ref{fig:lowVShigh}. As is clear, for $\beta_\epsilon$ below the regularization threshold (left panels), a critical transition towards the retrieval solution occurs, with the critical value of $\alpha$ depending on the training temperature $T_2$. A complete characterization of critical behavior of the RBM below the regularization threshold is reported in App. \ref{app:critical}. The generative capabilities are rather poor, since the feature magnetization $\tilde m$ is far from its maximum value $1$ even for a large relative width of the hidden layer. Further, for these choice of the tunable parameter $\beta_\epsilon$ and $T_2$, also spin-glass and ergodic solutions do exist; in all of our tests, as long as the order parameters are non-vanishing, the former turns out to be the one yielding minimal free energy, with ergodic and retrieval solutions being only local minima (in particular, the retrieval one is the one achieving highest free energy). Above the regularization threshold (right panel), the retrieval solution always exhibits a non-vanishing (and relatively high) value of the feature magnetization.

\section{Numerical results}\label{sec:numerical}
In this section, we present a compendium of numerical results for the training procedure of binary-Gaussian RBMs, focusing on the relaxation towards equilibrium of the learning dynamics and on the sampling capabilities after training. In all numerical experiments, the training is performed with the discretized Langevin equation
\begin{equation}
	 W_{i\mu}^{(k+1)}=W_{i\mu}^{(k)}-\delta \big(\langle \sigma_i s_\mu \rangle_+^{(k)}- \langle \sigma_i s_\mu \rangle_ -^{(k)}\big)-\delta  \frac{\eta N }{\beta_\epsilon}W_{i\mu}^{(k)}+\sqrt {2\delta T_2 }\eta_{i\mu}^{(k)},
\end{equation}
with $\eta_{i\mu}^{(k)} \sim\mathcal N(0,1)$ {i.i.d.} random variables. In our simulations, the learning rate is $\delta=10^{-3} $, the number of epochs (unless explicitly stated) is $S=10^4$, the dataset size is $M=10^4$, the size of the visible layer is $N_\sigma = 200$, and the positive and negative gradients (resp. $\langle\cdot \rangle_+^{(k)}$ and $\langle\cdot \rangle_-^{(k)}$) are estimated at each epoch with Persistent Contrastive Divergence algorithm (PCD) \cite{tieleman2008training}. In generative operational mode, visible layer's activity is prepared in a configuration $\bb \sigma^0$ and updated by the usual RBM 2-step procedure, e.g. $\bb\sigma^n\to \bb\sigma^{n+1}$ results from $\bb s^n \sim P(\bb s \vert \bb \sigma^n)$ and $\bb\sigma^{n+1} \sim P (\bb\sigma\vert \bb s^n)$, with the sampling temperature fixed to $T_1 =1$.

\subsection{Quantifying training performances}

As a first numerical check, we investigate the learning performances of the RBM during training. Because of the impracticability of computing the (empirical) KL divergence (due to the presence of the model partition function) in extensive numerical simulations, we follow a pseudo-(log)likelihood (likelihood from now on) $\mathcal L_D$ quantification of the training performances, see App. \ref{app:train_perfs}.
\begin{figure}
    \centering
    \includegraphics[width=\textwidth]{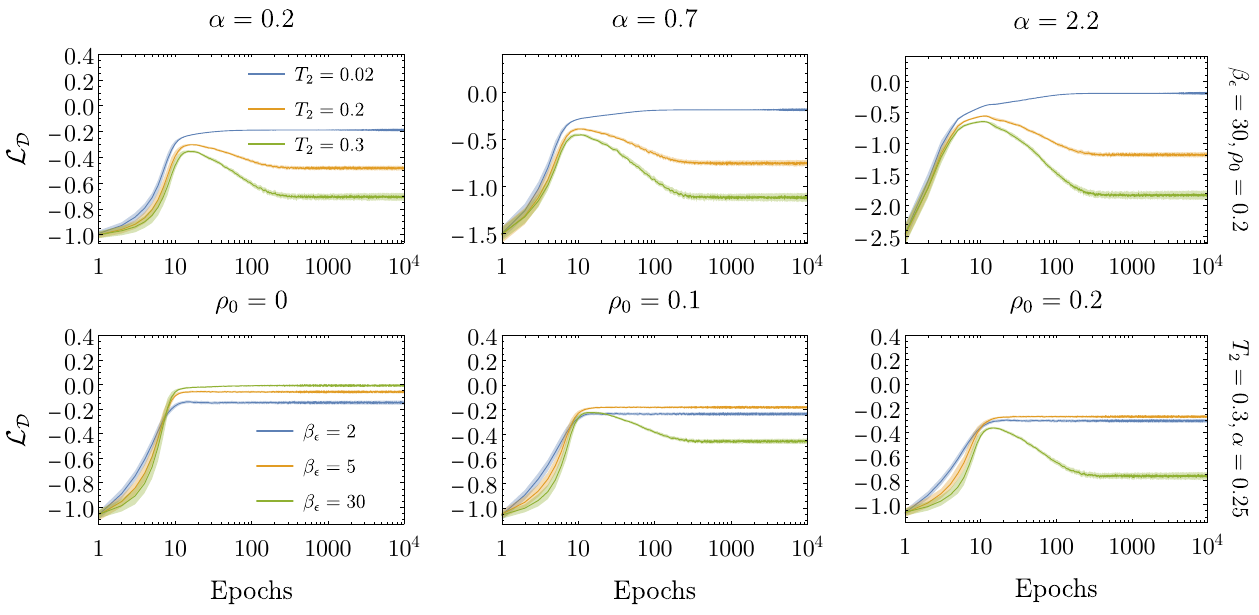}
    \caption{{\bfseries Training performances of the binary-Gaussian RBMs.} The figure shows the evolution of the likelihood function during the training procedure for various values of the external parameters. First row: we fix $\beta_\epsilon=30$ and $\rho_0=0.2$ and vary $\alpha=0.2,0.7,2.2$ (resp. left, middle and right plots) and $T_2=0.02,0.2,0.3$ (resp. blue, orange and green curves). Second row: we fix $T_2 =0.3$ and $\alpha=0.25$ and vary $\rho_0=0,0.1,0.2$ (resp. left, middle and right plots) and $\beta_\epsilon =2,5,30$ (resp. blue, orange and green curves). The curves are the average over 100 different realization of the training procedure for fixed external parameters, and the associated filled regions is the inter-quartile range at each training epoch.}
    \label{fig:training_performances}
\end{figure}
Numerical outcomes are reported in Fig. \ref{fig:training_performances}. In the first row, we reported the evolution of the likelihood function during training at $\beta_\epsilon=30$ and $\rho_0=0.2$, for various values of the relative size $\alpha$ of the hidden layer ($\alpha=0.2$ on the left, $\alpha=0.7$ in the center plot and $\alpha=2.2$ on the right), for various values of $T_2$; conversely, in the second row, we fixed $T_2 =0.3$ and $\alpha=0.25$, and vary $\rho_0$ ($=0,0.1,0.2$ from the left to the right) and $\beta_\epsilon=2,5,30$. As is clear, the final results of the likelihood function crucially depend on the combination of these tunable parameters. In particular, increasing the temperature for non-zero $\rho_0$ yields a poor information-extraction power of the RBMs,\footnote{By inspecting at the plots, it is clear that early-stopping criterion can be adopted in such cases. However, here we only focus on the relaxation of the training procedure} resulting in a non-monotonous behavior of the likelihood as a function of the training time. A similar fate is met if we increase $\beta_\epsilon$ for non-zero $\rho_0$ and high $T_2$. It is instructive to compare these results with the response diagrams reported in Fig. \ref{fig:phase2}. First, let us notice that the non-monotonicity of the likelihood takes place close to or above the curve $T_2 = (1+\rho_0)/(\beta_\epsilon \rho_0)$ at which the RS solution is expected to fail (recall that for $\rho_0=0.2$ and $\beta_\epsilon=30$ -- i.e. first row -- the failure for the RS solution holds at $T_2=0.2$).\footnote{Due to the relatively small size of the networks here considered, the transition towards the non-RS regime does not hold abruptly.} Also, plots in the first row highlight that the qualitative behavior of the likelihood function is unchanged as $\alpha$ is increased, the main difference being the slightly higher final likelihood values at large $\alpha$. This is consistent with the fact that the transition towards non-RS regime is independent of $\alpha$, as we found in the response diagrams. On the other hand, far from pathological choices of the external parameters, the likelihood increases monotonously with the training time, and the information extraction capabilities improves as $\alpha$ is increased. These results signal that the response behavior shown in Fig. \ref{fig:phase2} is consistent with the information extraction performances of these RBMs.

\subsection{Test of RS solution in the generative regime}\label{subsec:check_RS}

\begin{figure}
    \centering
    \includegraphics[width=\textwidth]{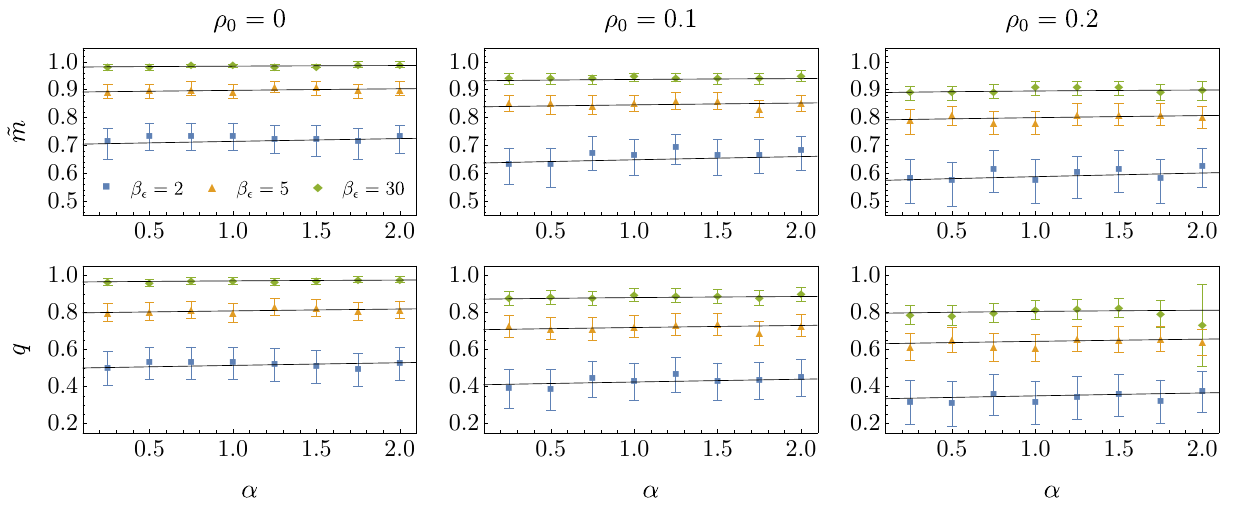}
    \caption{{\bfseries Numerical check of the RS solution at low training temperature $T_2$.} The figure shows the comparison between the theoretical predictions at low training temperature $T_2=0.02$ for various values of the remaining external parameters. In the first row, we reported the results for the feature magnetization $\tilde m$, while in the second one the same is shown for the visible layer overlap. The symbols represent experimental, computational findings, while the solid black curves represent the solution of the self-consistency equations \eqref{eq:n_final_0}-\eqref{eq:q_final_0} obtained under the RS ansatz.}
    \label{fig:RS_check}
\end{figure}

We now turn on checking the effectiveness of the RS solution far from the non-RS regime. To do this, we fix $T_2=0.02$ and vary the remaining external parameters. After the training procedure is over, we prepare the visible layer in a random configuration, and let the system evolve for $10^4$ updates of the whole network. We repeat the procedure for $10^4$ different initial conditions, and compute the feature magnetization $\tilde m$, then average over all the initial conditions and over 30 different realization of the training procedure. The results are reported in the first row of Fig. \ref{fig:RS_check}. We also computed the visible layer overlap as
\begin{equation}\label{eq:Wstar}
q(\bb W^*) = \frac1{N_\sigma} \sum_{i=1}^{N_\sigma} \langle \sigma_i \rangle_{\bb W^*}^2,
\end{equation}
where the $\langle \cdot \rangle_{\bb W^*}$ is the sample average for a fixed realization of the weights $\bb W^*$ resulting from the training procedure. In order to estimate its variance, we also computed the second moment of the overlap distribution, namely $q_2(\bb W^*)=\frac1{N_\sigma^2} \sum_{i,j=1}^{N_\sigma} \langle \sigma_i\sigma_j \rangle_{\bb W^*}^2$. The numerical results are then averaged over 30 different realization of the training procedure in the same setting of the external parameters. The results are reported in the second row of Fig. \ref{fig:RS_check}.
\par\medskip
The good agreement between the theoretical predictions and the numerical results (the only appreciable deviation concerns the overlap curves at high values of $\alpha$, $\rho_0$ and $\beta_\epsilon$, a case which we deepen in the following sections) suggests that our RS solution works well in describing the generative performances of the RBM at low training temperature $T_2$.

\subsection{Numerical evidences for RSB}\label{subsec:RSB}
\begin{figure}
    \centering
    \includegraphics[width=0.55\textwidth]{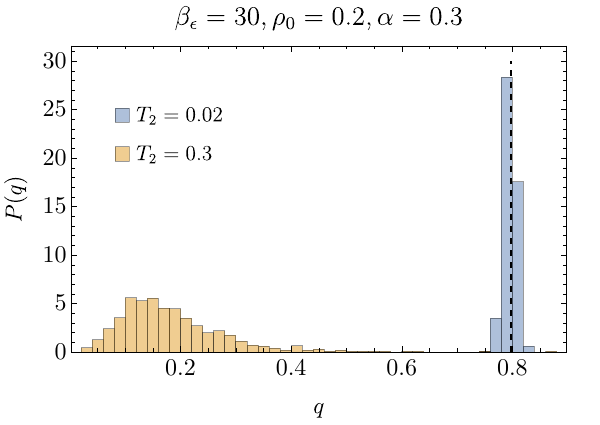}
    \caption{{\bfseries Empirical probability distribution of the overlaps.} The plot shows the overlap distributions at $\alpha=0.3$, $\beta_\epsilon=30$ and $\rho_0=0.2$ in the RS-regime ($T_2=0.02$) and the non-RS one ($T_2=0.3$). The dashed vertical line is the theoretical prediction given by the RS solution given by Eqs.~\eqref{eq:n_final_0}-\eqref{eq:q_final_0}, at $T_2=0.02$ (in the non-RS regime there is no theoretical prediction). The distributions are realized computing the overlap as in the previous section with 1000 different realizations of the training procedure.}
    \label{fig:overlap_dists}
\end{figure}

Once the effectiveness of the RS solution is ensured, the next step is testing the phenomenology of the system after training above the curve $T_2=(1+\rho_0)/(\beta_\epsilon \rho_0)$, i.e. the region where the RBMs are expected not to exhibit RS behavior. To this goal we inspect the overlap distribution: we fix $\beta_\epsilon =30$ and $\rho_0=0.2$, and then compare the overlap distributions at $T_2=0.02$ (RS regime) and $T_2=0.3$ (non-RS regime); in both cases we set $\alpha=0.3$ (we recall that the transition towards non-RS regime does not depend on the relative size of the hidden layer). The distributions are obtained computing the order parameter as defined in Eq.~\eqref{eq:Wstar}, collecting the results for 1000 different realizations of the training procedure; the results are reported in Fig. \ref{fig:overlap_dists}.
\par\medskip
In the RS regime ($T_2=0.02)$, the overlap distribution is peaked around the theoretical prediction (vertical dashed line), with a variance compatible with finite-size fluctuations around the expectation value. Pushing the system in the non-RS regime, the overlap distribution displays a totally different shape, exhibiting a wide continuous bulk at low values of $q$ with no isolated peaks. This suggests that a more complex phenomenology (w.r.t. the pure RS case) takes place in this region of the parameter space.
\par
\begin{figure}
    \centering
    \includegraphics[width=\textwidth]{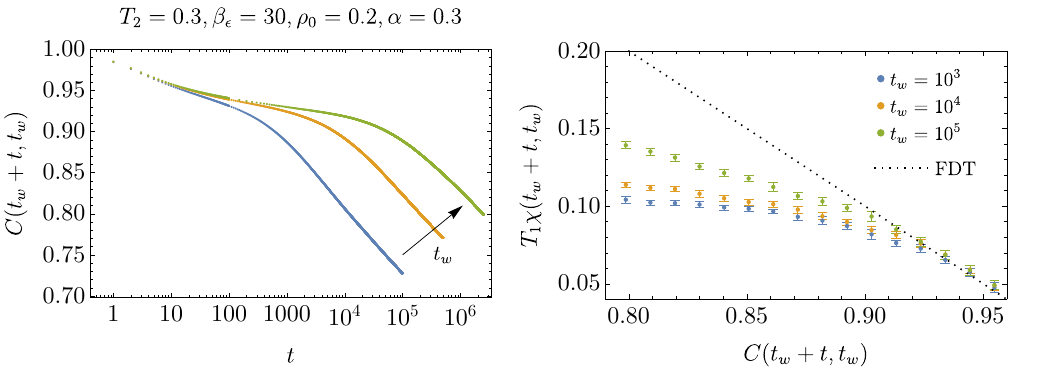}
    \caption{{\bfseries Aging and FDT violations in the non-RS regime.} The figure shows the numerical results in the non-equilibrium dynamics of the RBM in the non-RS regime. In particular, in the left plot we reported the dependence of the auto-correlation functions as a function of $t$ for various waiting times $t_w$. The numerical results are averaged over $2\cdot 10^4$ different thermal histories and $50$ different realizations of the training procedure. Further, for the $\chi \ vs \ C$ plot, we divide the horizontal axis in equally spaced intervals, and use the mean values of points falling in each of them both for the autocorrelation and the linear response functions (for the latter, also the standard deviation is shown).}
    \label{fig:FDT}
\end{figure}

\medskip
In order to better support the emergence of pure RSB behavior, we fix $T_2=0.3$ and $\alpha=0.3$, $\rho_0=0.2$, $\beta_\epsilon=30$, and study the dynamical violations of the fluctuation-dissipation theorem (FDT) \cite{kubo1966fluctuation}. In our setup, we adopted the zero-field cooling schedule \cite{marinari1998violation} and perform a (rough) version of the analysis reported in \cite{baity2017statics,paga2021spin}. Namely, we prepare the system in a random initial condition and let it evolve undisturbed for a time lapse $t_w$, whereupon a (homogeneous) field $h_i=h \Theta(t-t_w)$ is applied to the network. After the magnetic field is switched on, the (global) magnetization curve $m(t_w+t,t_w) = \frac1{N_\sigma}  \sum_{i=1}^{N_\sigma} \llangle \sigma_i(t+t_w)\rrangle$ is computed (with $\llangle \cdot \rrangle$ being an average over different thermal histories and realizations of the training procedure for a given choice of the external parameters). The magnetization curves are used to estimate the linear susceptibility as \cite{baity2017statics}:
$$
\frac{m(t_w+t,t_w)}{h}=\chi (t_w+t,t_w)-\frac{h^2}{3!}\chi_3 (t_w+t,t_w)-\frac{h^4}{5!}\chi_5 (t_w+t,t_w).
$$
Here, the non-linear susceptibilities $\chi_{3,5} (t_w+t,t_w)$, and $\mathcal O (h^5)$ corrections are neglected. Then, we perform the zero-field cooling schedule at three different values of the external magnetic field, i.e. $h=0.02,0.04,0.08$, and determine $\chi (t_w+t,t_w)$ (as well as the non-linear ones, although they are not used in the numerical simulations) by solving the resulting linear system. The auto-correlation function
$$
C(t_w+t,t_w) = \frac1{N_\sigma}  \sum_{i=1}^{N_\sigma}\llangle \sigma_i (t+t_w)\sigma_i (t_w)\rrangle,
$$
is also estimated numerically. The whole procedure is repeated for $t_w= 10^3,10^4,10^5$. After having collected these dynamical observables, we test the validity of the FDT, as shown in Fig.~\ref{fig:FDT} (see the caption for further details).
First, we notice that the the auto-correlation functions displays the typical behavior of aging systems \cite{cugliandolo1997energy,crisanti2003violation,ritort2003glassy,martin2023numerical}, as the behavior of the autocorrelation functions depend on the waiting time $t_w$, signaling that time-translation invariance does not hold. More remarkably, in the non-RS regime, the FDT is dynamically violated, and (even though our measurements allows the inspection of a limited portion of the $\chi\ vs\ C$ plane) the shape of the deviation from the baseline $\chi = 1-C$ (recall that sampling temperature is $T_1=1$) suggests the existence of multiple timescales in non-equilibrium dynamics, a signature of RSB effects \cite{crisanti2003violation,ritort2003glassy}.
\par
\begin{figure}
    \centering
    \includegraphics[width=\textwidth]{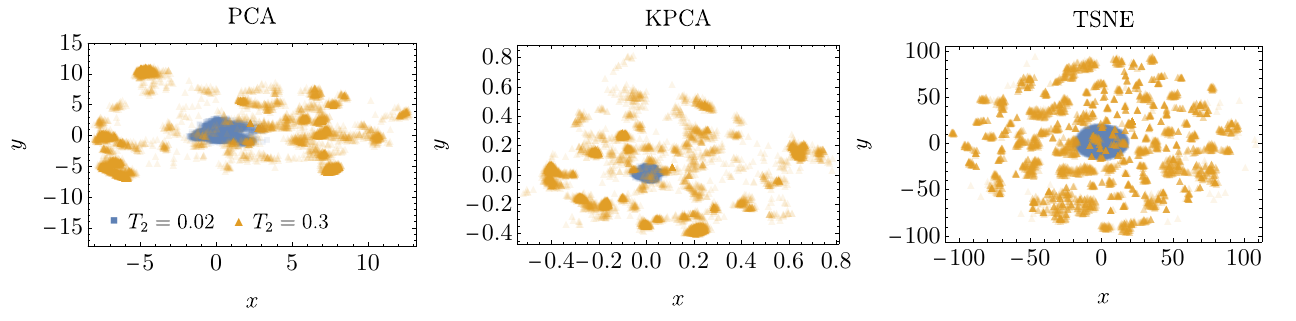}
    \caption{{\bfseries 2D visualizations of generated samples.} The figure show 2D visualizations of samples generated by the RBMs, and obtained by PCA (left), sigmoid Kernel-PCA (center) and TSNE (right). The external parameters are fixed as $\rho_0 =0.2$, $\beta_\epsilon =30$ and $\alpha=0.3$, while the training temperature is $T_2=0.02$ (blue points) and $T_2 =0.3$ (orange triangles). The samples are generated by a single realization of the training procedure.}
    \label{fig:2d_embedding}
\end{figure}

\medskip
As a final experiment on this point, we also perform a 2D dimensional representation of the samples generated by the machine. The experiment is the following: we take a single realization of the training procedure and prepare the system in a random initial configuration, then we let the system evolve for a waiting time $t_w=10^4$ updates of the whole network. After that, 20 samples are collected at a temporal distance of 100 updates (in the RS regime this choice ensures that the system completely loses the correlation with the previous sample, see next subsection). The procedure is repeated for 500 different random initial conditions. After the samples are generated, they are embedded in 2-dimensional space according to principal component analysis (PCA), sigmoid Kernel-PCA \cite{scholkopf1997kernel} and $t$-distributed stochastic-neighbor-embedding (TSNE) \cite{van2008visualizing} algorithms.\footnote{We stress that all of these methods are correlation/distance-based algorithms for high-dimensional data visualization, thus they carry (although generally in a non-trivial way) information about the overlap between the generated samples.} The results are reported in Fig. \ref{fig:2d_embedding}. We find that, for low $T_2$ (RS regime), the generated samples exhibit large correlation among themselves, so that they reduction results in a single giant cluster around the origin. Conversely, in the non-RS regime we obtain many different clusters which are mutually low-correlated, so that the dimensional reduction pushes them far from each other. This feature holds for all of the three methods considered, meaning that generated samples exhibit a complicated clustered structure: from an intuitive point of view, this can be understood by thinking of these samples as trapped in different valleys, and points in two distinct valleys exhibit low overlap w.r.t. two samples in the same cluster. This (na\"ive) interpretation, together with the results discussed so far in this section and the absence of RS solution above $T_2=(1+\rho_0)/(\beta_\epsilon \rho_0)$, suggest that the system is there characterized by a pure RSB behavior. Notice that the appearance of RSB at high $T_2$ could appear somewhat counter-intuitive, as breaking of replica-symmetry is typically met at low temperature. However, this issues is only apparent. Indeed, we stress that the relevant temperature in the generative regime, i.e. the sampling temperature $T_1$, is fixed to 1 in our experiment. Further, according to our previous discussion (see Sec. \ref{subsec:theo_cons}), $T_2$ sets a scale in the sampling temperature separating {\it low} and {\it high} thermal noise levels: increasing $T_2$ at fixed $T_1$ means that the range of low $T_1$ is enlarged, with the ultimate possibility to observe breaking of replica symmetry. Hence, RSB scheme is needed to unveil the physics in the non-RS regime.

\subsection{Empirical results on dynamics at high $\alpha$}
\begin{figure}[h!]
    \centering
    \begin{minipage}{0.49\textwidth}
         \includegraphics[width=\textwidth]{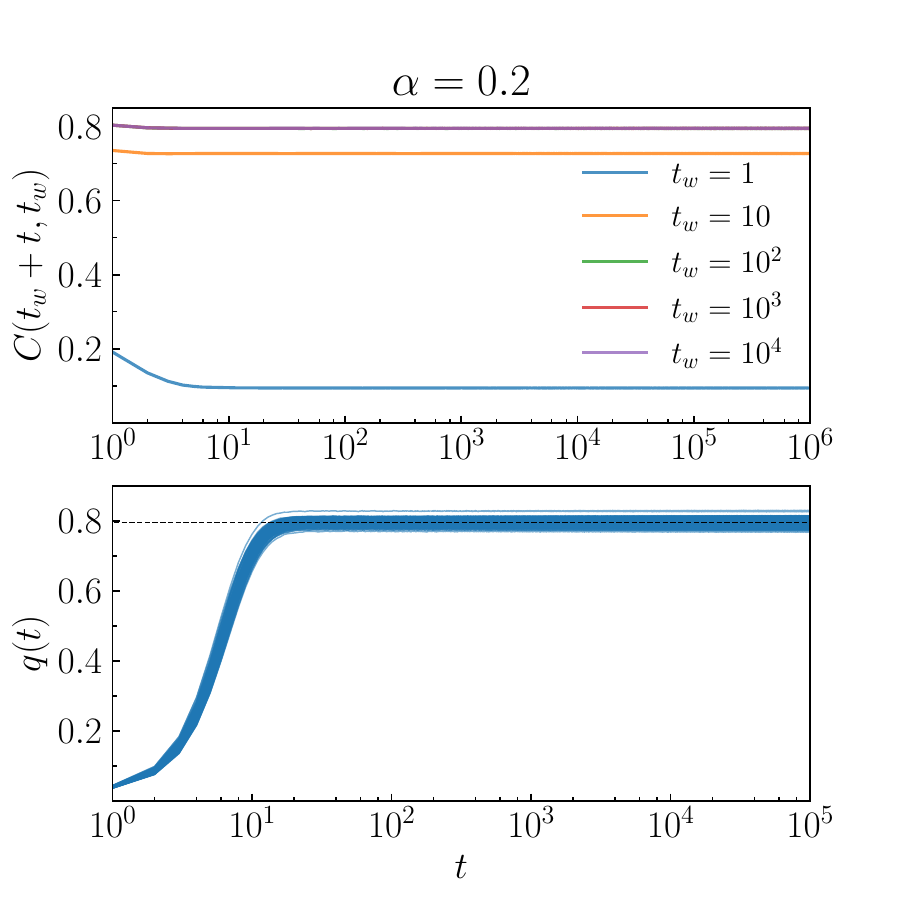}
    \end{minipage}
    \begin{minipage}{0.49\textwidth}
         \includegraphics[width=\textwidth]{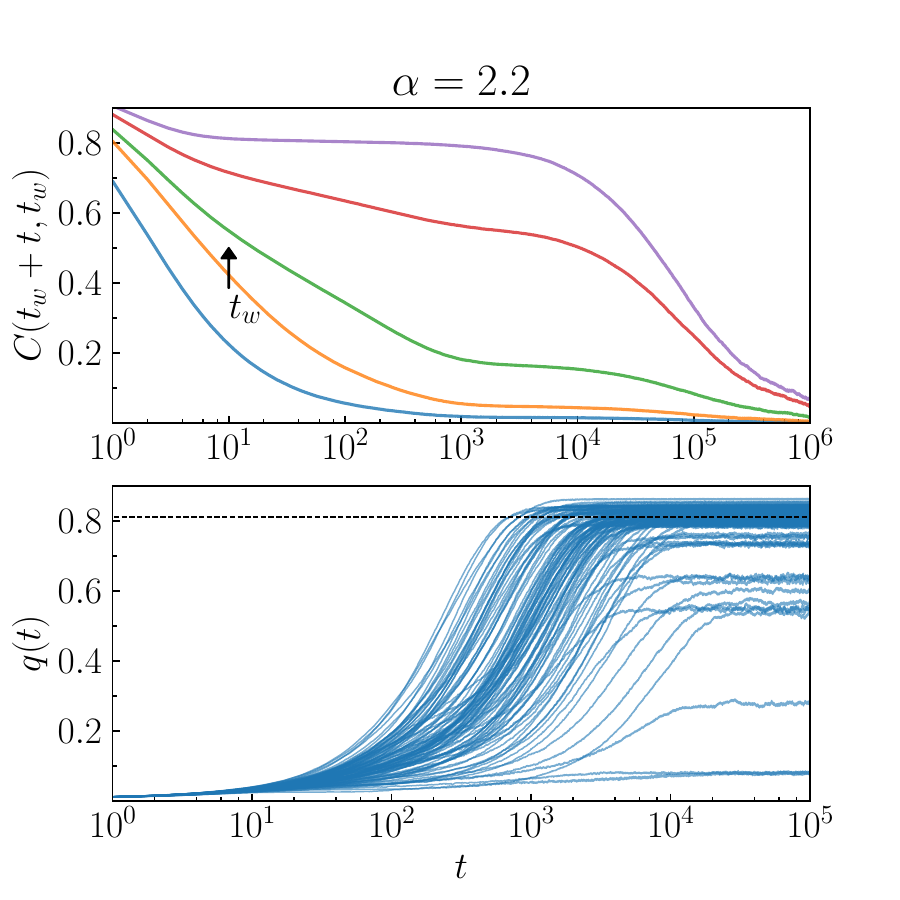}
    \end{minipage}
    \caption{{\bfseries Comparison of dynamical behavior for different $\alpha$ regimes.} The figure shows a compendium of the evolution of dynamical observables for low ($\alpha=0.2$, left column) and high ($\alpha=2.2$, right column) relative size of the hidden layer. First row: the auto-correlation function $C(t_w+t,t_w)$ is reported as a function of $t$ for different values of $t_w$($=1,10,10^2,10^3,10^4)$. The results have been carried out averaging over $10^4$ different thermal histories for each inferred model, and then averaged over $10$ different realization of the training procedures with fixed external parameters. Second row: the plots report the evolution of the overlap $q$ as a function of the sampling time $t$ starting from random initial conditions. At each time step, the overlap is computed by averaging over $10^4$ different samples. Each curve refers to 100 different realizations of the training procedure.}
    \label{fig:highalpha}
\end{figure}

As anticipated in Subsec. \ref{subsec:check_RS}, we now inspect the sampling dynamics of RBMs at low $T_2$ and high $\alpha$, $\rho_0$ and $\beta_\epsilon$, where deviation from the theoretical predictions of the overlap were observed. Again, we focus on the auto-correlation functions as well as the evolution of the overlap $q(t)$ (for different realizations of the training procedure) during the generative mode. We fix $T_2=0.02$ and compare $\alpha =0.2$ and $\alpha=2.2$. Numerical results are reported in Fig. \ref{fig:highalpha}.
At low $\alpha$, the system behaves nicely in the generative mode. In particular, regardless of the value of $t_w$, the overlap of the visible layer at time $t+t_w$ w.r.t. the reference configuration at time $t_w$ gradually reduces to a constant value after $\gtrsim10$ updates (we checked that the system accordingly loses memory of the reference configuration by computing the auto-covariance function $\hat C(t+t_w,t_w)=C(t+t_w,t_w)-\frac1{N_\sigma} \sum_i \llangle \sigma_i (t_w+t)\rrangle\llangle \sigma_i (t_w)\rrangle$, going to zero for $t\gtrsim 10$). In particular, it is clear that $\lim_{t\to\infty}\lim_{t_w\to\infty} C(t+t_w,t_w)=q_{RS}$ as expected, with $q_{RS}$ being the theoretical prediction as given by the RS solution (see the black dashed line in the lower left plot). This is also clear from the temporal dependence of $q$: preparing the system in random configurations, after few update steps the generated sample exhibit the overlap predicted by the RS theory. On the other hand, increasing $\alpha$ results in a more complex phenomenology, as the auto-correlation function displays a non-trivial dependence both on $t$ and $t_w$. The temporal evolution of the overlap mirrors this behavior: in this regime, most of the training procedures will result in models generating samples with desired statistics, that is $q\sim q_{RS}$ (although with a longer relaxation time w.r.t. the low $\alpha$ case); however, even after $10^5$ updates of the network, the probability of obtaining trained machines that fail in generating good samples is non-vanishing and, in particular, one finds different values of the overlap between generated samples. This signals that, even if RS solution is present, the equilibrium dynamics could be rich. Understanding the origin of this abrupt change in the phenomenology would require a separate analysis (both from the statics -- i.e. again addressing RSB scheme and compare the associated solution with the RS approximation -- and dynamical point of view) going beyond the scope of this paper. 

\section{Conclusions and further developments} \label{sec:outro}
The development of a solid theory encompassing paradigmatic models in Machine Learning -- as, in the present case, the RBMs -- allows us to improve our understanding of the information-extraction mechanisms occurring during training and of the sampling capabilities of generative models, possibly contributing to the design of sustainable and interpretable devices and algorithms.
\newline
As emphasized in the introduction, this can be addressed from several point of views and, in the present paper, we consider the simple setting of a binary-Gaussian RBM inferring a single feature (or { direction}) from a noisy set of data, which can be viewed -- due to the simplicity of the problem under consideration -- as a minimal problem for studying intrinsic functioning principles of these generative models. We analyzed it by means of statistical mechanics of spin-glasses in the operative mode, with the quenched disorder being fixed by the posterior distribution resulting from the minimization of the empirical Kullback-Leibler divergence. Remarkably, despite the simplicity of the problem at our hand, the emerging phenomenology turns out to be very rich. In this scenario, we gave a complete (theoretical) characterization of the RS regime, and precisely identified working regimes in the hyper-parameters space. In particular, we identified a regularization threshold above which the system works properly (meaning that RS spin-glass and ergodic solutions are not present). Our theoretical findings further suggest the presence of a transition in the training temperature, above which the system is expected to exhibit pure RSB behavior. This conjecture is supported by numerical simulations, as both aging phenomena and FDT violations are observed. Despite all the results here reported, crucial points are left open for future investigations. The first theoretical challenge will consists in the application of RSB scheme, which would allow us to explore physics in the high training-temperature regime (and possibly at high relative size of the hidden layer), and thus clarify the origin of the pure RSB effects. This would constitute a significant contribution for addressing interpretability of information processing principles in generative models. Dynamical analysis would also be an interesting extension of the present work. Besides the technical aspects, a natural extension would be to go beyond the restricted nature of these models, for instance analyzing the role of horizontal connections \cite{fachechi2022outperforming,shi2019new}, as well as to inspect different type of units (e.g. the binary-binary RBM) and more realistic choices for the structured dataset, as recently tackled in \cite{fernandez2024replica}, focusing in particular on relevant scalings between the size and the number of training parameters, and thus addressing the emergence of overfitting phenomenon.

\section{Technical details}
Numerical simulations have been performed with TensorFlow 2.11 with CUDA Toolkit 11.7 and CuDNN 8.5 on a NVidia GeForce RTX-4070 Ti.

\section{Acknowledgments}
We acknowledge financial support from PNRR MUR project PE0000013-FAIR and from Sapienza University of Rome (RM120172B8066CB0, AR2221815D7192C1, AR1221815EA97525).  
\newline
This work has been supported by The Alan Turing Institute through the Theory and Methods Challenge Fortnights event "Physics-informed machine learning", which took place on 16-27 January 2023 at The Alan Turing Institute headquarters.
\newline
The Authors are very grateful to Prof. A. Barra for fruitful discussions.

%%%%%%%%%%%%%%%%%%%%%%%%

\appendix

\section{On the Machine Learning interpretation of $T_2$}\label{app:T2}

In realistic applications of gradient descent, stochastic noise arises due to the fact that averages of the positive and negative gradients are performed over a mini-batch (MB) rather than on the whole dataset (the batch B). In this appendix, we examine the relation between the temperature $T_2$ and the size of mini-batches giving rise to the same amount of noise at the very early stages of the training: transitions in $T_2$ should thus be intended as transitions w.r.t. the mini-batch size, see also \cite{marino2024phase}. For the sake of clearness, we make explicit the role of the training temperature $T_2$ in the Gradient Flow equation \eqref{eq:stochastic1} by rescaling the stochastic process (we neglect the $L_2$-regularization without loss of generality), so that
\begin{equation}
	\frac{d W_{i\mu}}{dt} = -\frac{\partial D_{KL}}{\partial W_{i\mu}}+\sqrt{2T_2} \eta_{i\mu} (t),
\end{equation}
where now $\mathbb{E}_{\eta}\eta_{i\mu}(t) \eta_{j\nu} (t') = \delta_{ij} \delta_{\mu\nu} \delta (t-t')$. Let us now consider the two approaches, {\it i.e.} the (batch) gradient descent with external Gaussian noise and its mini-batch version (at $T_2=0$), namely
\begin{eqnarray}
	\frac{d W_{i\mu}}{dt}& =& -\frac{\partial D_{KL}}{\partial W_{i\mu}}\Big\vert_{\textrm{B}}+\sqrt{2T_2} \eta_{i\mu} (t),\\
		\frac{d W_{i\mu}}{dt}& =& -\frac{\partial D_{KL}}{\partial W_{i\mu}}\Big\vert_{\textrm{MB}},
\end{eqnarray}
Let us stress that the dependence on the dataset size passes only for clamped correlation function $\langle  \cdot \rangle_{\bb W, +}$, the negative gradients being in common in the two approaches. Let us now estimate the clamped correlation function. In any case
\begin{equation}
	\langle \sigma_i s_\mu \rangle_{\bb W, +} \approx \frac1D\sum_{A=1}^D \bar \sigma_i ^A \int d\bb s \, P_{\bb W} (\bb s \vert \bar {\bb \sigma}^A)\bar \sigma_i ^A s_\mu,
\end{equation}
where $D$ is the size of the samples, namely $D=M$ for the batch case and $D=m\ll M$ for the mini-batch scenario. In our scenario, the conditioned expectation value is $\int d\bb s P _{\bb W}(\bb s \vert \bar {\bb \sigma}^A)s_\mu = \sum_j W_{j\mu}\bar \sigma_j ^A$, so that
\begin{equation}
	\langle \sigma_i s_\mu \rangle_{\bb W, +}\approx  \frac1D\sum_{A=1}^D \sum_j W_{j\mu}\bar \sigma_i ^A \bar \sigma_j ^A= W_{i\mu}+ \sum_{j\neq i } W_{j\mu }\frac 1D \sum_{A=1}^D \bar \sigma_i ^A \bar \sigma_j ^A.
\end{equation}
Let us now consider the trajectories $\bb W^{\textrm{B}}$ and $\bb W^{\textrm{MB}}$ of the weights under the gradient flow in the two cases starting from the same initial condition, say, for instance, $W_{i\mu}^0{\underset{i.i.d.} \sim }\mathcal N(0,N^{-1}).$ Thus, setting again $\beta_1=1$ for convenience, we have
\begin{eqnarray}
    \frac{dW^{\textrm{B}}_{i\mu}}{dt}&=& W^{\textrm{B}}_{i\mu}+ \sum_{j\neq i } W^{\textrm{B}}_{j\mu }\frac 1M \sum_{A=1}^M \bar \sigma_i ^A \bar \sigma_j ^A-\langle \sigma_i s_\mu \rangle_{\bb W^{\textrm{B}},-}+\sqrt{2T_2}\eta_{i\mu}(t),\\
    \frac{dW^{\textrm{MB}}_{i\mu}}{dt}&=& W_{i\mu}^{\textrm{MB}}+ \sum_{j\neq i } W_{j\mu }^{\textrm{MB}}\frac 1m \sum_{s=1}^m \bar \sigma_i ^{A_s }\bar \sigma_j ^{A_s}-\langle \sigma_i s_\mu \rangle_{\bb W^{\textrm{MB}},-}.
\end{eqnarray}
At the very early stages of training, it is reasonable to assume that the configuration of the weights is close to the initial condition, so that $W_{i\mu}^{\textrm{B}}\approx W_{i\mu}^{\textrm{MB}}\approx W_{i\mu}^{0}$ and $\langle \sigma_i s_\mu \rangle_{\bb W^{\textrm{B}},-}\approx \langle \sigma_i s_\mu \rangle_{\bb W^{\textrm{MB}},-}\approx \langle \sigma_i s_\mu \rangle_{\bb W^{0},-}$. Further, in this case we can also assume that the weights are weakly correlated with each other and with the empirical data. Under these assumptions, we can approximate the second contributions by applying the central limit theorem. In particular, one can show that $\mE (W_{j\mu }^0\frac 1D \sum_{A=1}^D \bar \sigma_i ^A \bar \sigma_j ^A)=0$ (where $\mathbb{E}$ is the average taken over all possible initial conditions of the weights and all possible realizations of the dataset) and $\textrm{Var}(W_{j\mu }^0\frac 1D \sum_{A=1}^D \bar \sigma_i ^A \bar \sigma_j ^A)=\frac1N\big(\frac{1-r^4}{D}+r^4\big)$, thus
\begin{eqnarray}
    \frac{dW^{\textrm{B}}_{i\mu}}{dt}&\sim& W^{0}_{i\mu}-\langle \sigma_i s_\mu \rangle_{\bb W^{0},-}+ \sqrt{\frac{1-r^4}{M}+r^4}z_{i\mu}+\sqrt{2T_2}\eta_{i\mu}(t),\\
    \frac{dW^{\textrm{MB}}_{i\mu}}{dt}&\sim& W^{0}_{i\mu}-\langle \sigma_i s_\mu \rangle_{\bb W^{0},-}+ \sqrt{\frac{1-r^4}{m}+r^4}z'_{i\mu},
\end{eqnarray}
with $z_{i\mu},z'_{i\mu}\underset{i.i.d.}\sim\mathcal N(0,1)$.
In order for the two version to exhibit the same amount of noise, we require
$$
 \Big(\frac{1-r^4}{M}+r^4\Big)+2T_2 =\Big(\frac{1-r^4}{m}+r^4\Big),
$$
and, solving for $T_2$, we get
\begin{equation}
	T_2 =  \frac{1-r^4}{2}\Big(\frac1m -\frac1M\Big)= \frac{r^2(1+r^2)}{2} (\rho _m -\rho),
\end{equation}
where $\rho =(1-r^2)/Mr^2$ is the dataset entropy and $\rho_m = (1-r^2)/mr^2 $ is the analogous quantity for the minibatch. The same results also hold for the regularized case (as it is a common contribution for both the approaches). Clearly, in machine-learning applications one has minibatches of at least hundreds of examples (to make the positive and negative gradients properly estimated), so that realistic values of the training temperature would be $T_2\sim 10^{-2}$. Alternatively, one can train the network via discretized Langevin equation, in order to have full control on stochasticity contributions in the Gradient Flow, and thus explore a wider range of temperatures.

\section{Details of replica computations}\label{app:details_replica}
In this Appendix, we will give the details of replica trick computations for the quantity
$$
\mE_{ \bb \chi}\Big[\int d\bb W \exp\big(-\frac{\beta_2N\epsilon_W}{2}\sum_{i\mu} W_{i\mu}^2\big)  Z_{\bb W}^{n_1}\prod_{A=1}^M Z_{\bb W}(\bar{\bb\sigma}^A)^{n_2}\Big]^{n_3},
$$
appearing in the quenched free energy \eqref{eq:tot_fe}. We recall that we set $\beta_1=1$ without loss of generality as it can be recovered {\it a posteriori} by rescaling the regularization parameter as $\epsilon_W\to \epsilon_W/\beta_1^2$. Let us start with the first factor:
\begin{equation}
	\begin{split}
		Z_{\bb W}^{n_1}&=\sum_{\bb\sigma^1\dots \bb\sigma^{n_1}}\sum_{\bb s^1\dots \bb s^{n_1}}\exp\Big(\sum_{a=1}^{n_1}\sum_{i\mu}W_{i\mu}\sigma_i ^a s_\mu^a
		\Big).
	\end{split}
\end{equation}
For the second one, we have
\begin{equation}
	\begin{split}
		\prod_{A=1}^M Z_{\bb W}(\bar{\bb\sigma}^A )^{n_2}&=\sum_{\bbb s^{(1,1)}\dots \bbb s^{(M, n_2)}}\exp\Big( \sum_{A=1}^M \sum_{\bar a=1}^{n_2}\sum_{i\mu}W_{i\mu}\bar \sigma_i ^A \bar s^{(A,\bar a)}_\mu
		\Big).
	\end{split}
\end{equation}
Putting the two pieces together, we get
\begin{equation*}
	\begin{split}
			Z_{\bb W}^{n_1}&	\prod_{A=1}^M Z_{\bb W}(\bar{\bb\sigma}^A )^{n_2}=\sum_{\bb\sigma^1\dots \bb\sigma^{n_1}}\sum_{\bb s^1\dots \bb s^{n_1}}\sum_{\bbb s^{(1,1)}\dots \bbb s^{(M ,n_2)}}\exp\Big[
			\sum_{i\mu} W_{i\mu}\big(\sum_{a=1}^{n_1}\sigma_i ^a s_\mu ^a+\sum_{A=1}^M \sum_{\bar a=1}^{n_2}\bar \sigma_i ^A \bar s_\mu ^{(A,\bar a)}\big)
			\Big].
	\end{split}
\end{equation*}
We now perform the Gaussian integration over the network parameters:
\begin{equation*}
	\begin{split}
		\int &	d\bb W \exp\big(-\frac{\beta_2N\epsilon_W}{2}\sum_{i\mu} W_{i\mu}^2\big) Z_{\bb W}^{n_1}\prod_{A=1}^M Z_{\bb W}(\bar{\bb\sigma}^A )^{n_2}=\\
		&=\frac1{\mathcal V}\sum_{\bb\sigma^1\dots \bb\sigma^{n_1}}\sum_{\bb s^1\dots \bb s^{n_1}}\sum_{\bbb s^{(1,1)}\dots \bbb s^{(M ,n_2)}}\exp\Big[
		\frac1{2N \beta_2 \epsilon_W}\sum_{i\mu} \big(\sum_{a=1}^{n_1}\sigma_i ^a s_\mu ^a+\sum_{A=1}^M \sum_{\bar a=1}^{n_2}\bar \sigma_i ^A \bar s_\mu ^{(A,\bar a)}\big)^2
		\Big],
	\end{split}
\end{equation*}
with $\mathcal V$ being unessential volume factor coming from the Gaussian integration. Finally, we perform the third replication, which leads to
\begin{equation*}
	\begin{split}
		&\Big[\int d\bb W \exp\big(-\frac{\beta_2N\epsilon_W}{2}\sum_{i\mu} W_{i\mu}^2\big) Z_{\bb W}^{n_1}	\prod_{A=1}^M Z_{\bb W}(\bar{\bb\sigma}^A )^{n_2}\Big]^{n_3}=\\
		&=\frac1{\mathcal V}\sum_{\bb\sigma^{1,1}\dots \bb\sigma^{n_1,n_3}}\sum_{\bb s^{1,1}\dots \bb s^{n_1,n_3}}\sum_{\bbb s^{(1,1),1}\dots \bbb s^{(M ,n_2),n_3}}\exp\Big[
		\frac1{2N \beta_2 \epsilon_W}\sum_{i\mu}\sum_{\gamma=1}^{n_3}\big(\sum_{a=1}^{n_1}\sigma_i ^{a\gamma} s_\mu ^{a\gamma}+\sum_{A=1}^M \sum_{\bar a=1}^{n_2}\bar \sigma_i ^A \bar s_\mu ^{(A,\bar a)\gamma}\big)^2
		\Big]=\\
		&
		\begin{split}
			=\frac1{\mathcal V}\sum_{\bb \sigma, \bb s, \bbb s}\exp\Big[\frac1{2N \beta _2 \epsilon_W}\sum_{i\mu\gamma}\big(\sum_{ab=1}^{n_1}\sigma^{a\gamma}_i \sigma^{b\gamma} _i s^{a\gamma}_\mu s^{b\gamma}_\mu&+2\sum_{a=1}^{n_1}\sum_{A=1}^M \sum_{\bar a = 1}^{n_2}\sigma_i ^{a\gamma}\bar \sigma_i ^A s^{a\gamma}_\mu \bar s^{(A\bar a)\gamma}_\mu+\sum_{AB=1}^M \sum_{\bar a \bar b =1 }^{n_2}
			\bar \sigma_i ^A \bar \sigma_i ^B \bar s_\mu ^{(A,\bar a)\gamma} \bar s_\mu ^{(B,\bar b)\gamma}
			\big)\Big],
		\end{split}
		\end{split}
\end{equation*}
where $\sum_{\bb\sigma,\bb s, \bbb s}$ is the sum over all types of replicas. Introducing the order parameters \eqref{eq:OP_visible}-\eqref{eq:OP_hidden}, we can write
\begin{equation}\nonumber
	\begin{split}
	&\mE _{\bb\chi}	\Big[		\int	d\bb W \exp\big(-\frac{\beta_2N\epsilon_W}{2}\sum_{i\mu} W_{i\mu}^2\big) Z_{\bb W}^{n_1}	\prod_{A=1}^M Z_{\bb W}(\bar{\bb\sigma}^A )^{n_2}\Big]^{n_3}=\\&=  \frac1{\mathcal V}\mE _{\bb\chi}\sum_{\bb\sigma, \bb s , \bbb s} \int\Big[\prod_{ab\gamma}dq_{ab}^\gamma dp_{ab}^\gamma \delta\big(q_{ab}^\gamma-\frac1{N_\sigma}\sum_i \sigma_i ^{a\gamma} \sigma_i ^{b\gamma}\big)\delta\big(p_{ab}^{\gamma}-\frac1{N_s}\sum_i \sigma_i ^{a\gamma} \sigma_i ^{b\gamma}\big)\Big]\\&\cdot
		\Big[\prod_{AB\bar a \bar b \gamma}d\bar p^\gamma_{(A\bar a)(B\bar b)}\delta\big(\bar p^\gamma_{(A\bar a)(B\bar b)}-\frac1{N_s} \sum_{\mu} \bar s_\mu^{(A \bar a)\gamma} \bar s_\mu ^{(B\bar b)\gamma}\big)\Big]
	\Big[\prod_{aA\bar a\gamma}d n_{a(A\bar a)}^\gamma\delta\big(n_{a(A\bar a)}^\gamma-\frac1{N_s} \sum_\mu s_\mu^{(A \bar a)\gamma} \bar s_\mu^{(B\bar b)\gamma}\big)\Big]\\&\cdot\Big[\prod_{A  a\gamma} dm_{aA}^\gamma\delta \big(m_{aA}^\gamma -\frac1{N_\sigma} \sum_i \sigma_i ^{a\gamma}\bar \sigma_i ^A\big)\Big] \exp\Big[\frac{{\eta (1-\eta)N}}{2\beta _2  \epsilon_W}\sum_\gamma\big(\sum_{ab}q^\gamma _{ab}p^\gamma_{ab}+2\sum_{aA\bar a}m^\gamma_{aA} n^{\gamma}_{a(A\bar a)}+\sum_{AB\bar a \bar b} q^E_{AB}\bar p ^\gamma_{(A\bar a)(B \bar b)}
		\big)\Big].
	\end{split}
\end{equation}
Using the Fourier representation of the Dirac delta, performing a Wick rotation on the emerging conjugate order parameters and suitably deforming the associated integration contour, we can rewrite the previous equation (upon absorbing all constants in the volume factor) as
\begin{equation}\label{eq:computations_1}
	\begin{split}
		&\mE _{\bb\chi}	\Big[	\int	d\bb W \exp\big(-\frac{\beta_2N\epsilon_W}{2}\sum_{i\mu} W_{i\mu}^2\big) Z_{\bb W}^{n_1}	\prod_{A=1}^M Z_{\bb W}(\bar{\bb\sigma}^A  )^{n_2}\Big]^{n_3}=\\&
=
		\frac1{\mathcal V}\mE_{\bb \chi}\int d \bb O d{\hat {\bb O}}\exp\Big(-N K(\bb O, \hat{\bb O})+N V(\bb O)\Big)\\
		& \cdot\sum_{\bb\sigma} \exp\Big[
		\frac1\eta \sum_{\gamma a A i} \hat m ^\gamma _{aA}\sigma_i ^{a\gamma}\bar \sigma_i ^A+\frac1\eta \sum_{\gamma ab i}\hat q^{\gamma }_{ab}\sigma_i ^{a\gamma}\sigma_i^{b\gamma}
		\Big]
		 \sum_{\bbb s}\exp \Big(\frac{1}{1-\eta}\sum_{\gamma AB \bar a \bar b\mu}\hat {\bar p}^\gamma_{(A\bar a)(B \bar b)}\bar s^{(A\bar a)\gamma}_\mu \bar s^{(B\bar b)\gamma}_\mu
		\Big)\\&\cdot
		\sum_{\bb s}\exp\Big(\frac1{1-\eta}\sum_{\gamma a A \bar a \mu}\hat n ^\gamma_{a(A \bar a)}s ^{a\gamma}_\mu \bar s^{(A\bar a)\gamma}_\mu+\frac1{1-\eta}\sum_{\gamma a b \mu}\hat p _{ab }^\gamma s^{a\gamma}_\mu s^{b\gamma}_\mu\Big),
	\end{split}
\end{equation}
where
\begin{eqnarray*}
\boldsymbol O&=&\{q_{ab}^\gamma,q_{ab}^\gamma,m_{aA}^{\gamma},n_{a(A\bar a)}^{\gamma},\bar p_{(A\bar a)(B\bar b)}^\gamma\},
\\
	K(\bb O, \hat{\bb O})&=&\sum_{\gamma ab} \hat q^\gamma _{ab}q^{\gamma}_{ab}
	+\sum_{\gamma ab} \hat p^\gamma _{ab}p^{\gamma}_{ab}+ \sum_{\gamma a A} \hat m^{\gamma}_{aA}m_{aA}^\gamma+\sum_{\gamma a A \bar a} \hat n^{\gamma}_{a(A\bar a)}n^\gamma_{a(A\bar a)}+ \sum_{\gamma AB \bar a \bar b} \hat {\bar p}^\gamma_{(A\bar a) (B\bar b)} {\bar p}^\gamma_{(A\bar a) (B\bar b)},
	\\
	V(\bb O)&=&\frac{{\eta (1-\eta)N}}{2\beta _2  \epsilon_W}\sum_\gamma\big(\sum_{ab}q^\gamma _{ab}p^\gamma_{ab}+2\sum_{aA\bar a}m^\gamma_{aA} n^{\gamma}_{a(A\bar a)}
	+\sum_{AB\bar a \bar b} q^E_{AB}\bar p ^\gamma_{(A\bar a)(B \bar b)}
	\big).
\end{eqnarray*}
The function $K$ is nothing but the scalar product of the order parameters with the associated conjugate fields, while the function $V$ specifies the interaction between the order parameters, and it strictly depends on the network architecture. We can now perform the Gaussian integration over the (clamped and free) hidden replicated units. {Starting with the contribution in the fourth line of Eq.~\eqref{eq:computations_1}}
\begin{equation}\label{eq:int_free_replicas}
	\begin{split}
		&\sum_{\bb s}\exp\Big(\frac1{1-\eta}\sum_{\gamma a A \bar a \mu}\hat n ^\gamma_{a(A \bar a)}s ^{a\gamma}_\mu \bar s^{(A\bar a)\gamma}_\mu+\frac1{1-\eta}\sum_{\gamma a b \mu}\hat p _{ab }^\gamma s^{a\gamma}_\mu s^{b\gamma}_\mu\Big)=
		\\
		=&	\prod_{\gamma \mu}\int \Big(\prod _a \frac{ds_\mu^{a\gamma}}{\sqrt{2\pi}}\Big)\exp\Big[-\frac12 \sum_{ab}s^{a\gamma}_\mu s^{b\gamma}_\mu\big(\delta_{ab}-\frac{2}{1-\eta}\hat p^\gamma_{ab}\big)+\frac1{1-\eta}\sum_a s^{a\gamma}_\mu\big(\sum_{A\bar a}\hat n ^\gamma_{a(A\bar a)} \bar s^{(A\bar a)\gamma}_\mu\big)
		\Big]=\\
		=&\prod_{\mu \gamma} \text{det}^{-1/2 }\big(\bb 1-\frac2{1-\eta}\hat{\bb p}^\gamma\big)
		\exp\Big[\frac1{2(1-\eta)^2}\sum_{ab}\sum_{AB\bar a \bar b} \hat n^\gamma_{a(A\bar a)}\bar s^{(A\bar a)\gamma}_\mu\big(\bb 1-\frac2{1-\eta}\hat{\bb p}^\gamma\big)^{-1}_{ab}\hat n^\gamma_{b(B\bar b)}\bar s^{(B\bar b)\gamma}_\mu\Big]=
		\\
		=&\exp\Big\{-\frac{1-\eta}{2}N \sum_\gamma \log \text{det}\big(\bb 1-\frac2{1-\eta}\hat{\bb p}^\gamma\big)+\frac1{2(1-\eta)^2}\sum_{\gamma AB\bar a \bar b\mu} \big[\sum_{ab}\hat n^\gamma_{a(A\bar a)}\big(\bb 1-\frac2{1-\eta}\hat{\bb p}^\gamma\big)^{-1}_{ab}\hat n^\gamma_{b(B\bar b)}\big]\bar s^{(A\bar a)\gamma}_\mu\bar s^{(B\bar b)\gamma}_\mu
		\Big\}.
	\end{split}
\end{equation}
{Here, the quadratic term in the clamped hidden units $\bar{\bb s}$ has the same structure of the last contribution in second line of Eq. \eqref{eq:computations_1}, so that it renormalizes the conjugate order parameter $\hat{\bar{\bb p}}$. Thus, defining the effective overlap}
\begin{equation}\label{eq:r_conj}
		r^\gamma_{(A\bar a)(B\bar b)}=	\hat{\bar p}^\gamma_{(A\bar a)(B\bar b)}+\frac1{2(1-\eta)}\sum_{ab} \hat n ^\gamma_{a(A\bar a)} \big(\bb 1 - \frac2{1-\eta}\hat{\bb p}^\gamma\big)^{-1}_{ab}\hat n ^\gamma_{b(B\bar b)},
\end{equation}
the sector of \eqref{eq:computations_1} involving clamped hidden {units can be computed in closed form as}
\begin{equation}
	\begin{split}
		&\sum_{\bbb s}\exp \Big(\frac{1}{1-\eta}\sum_{\gamma AB \bar a \bar b\mu}r^\gamma_{(A\bar a)(B \bar b)}\bar s^{(A\bar a)\gamma}_\mu \bar s^{(B\bar b)\gamma}_\mu
		\Big)=
		\\=&
		\prod_{\gamma\mu}\int \Big(\prod_{A\bar a} \frac{d\bar s^{(A\bar a)\gamma}_\mu}{\sqrt{2\pi}}\Big)\exp\Big[-\frac12 \sum_{AB\bar a \bar b} \bar s_\mu ^{(A\bar a)\gamma} \bar s_\mu^{(B\bar b)\gamma}\big(\delta_{AB}\delta_{\bar a \bar b}-\frac2{1-\eta}r^\gamma_{(A\bar a) (B\bar b)}\big)\Big]=\\
		=&\prod _{\gamma\mu}\det^{-1/2}_{Mn_2}\big(\bb 1-\frac2{1-\eta}\bb r^\gamma\big)=\exp\Big[-\frac{1-\eta}{2}N \sum_\gamma \log \det\big(\bb 1-\frac2{1-\eta}\bb r^\gamma\big) \Big]
		\end{split}
\end{equation}
Plugging these results in the expression \eqref{eq:computations_1}, we thus get
\begin{equation}\label{lastterm}
	\begin{split}
		&\mE _{\bb\chi}	\Big[\int	d\bb W \exp\big(-\frac{\beta_2N\epsilon_W}{2}\sum_{i\mu} W_{i\mu}^2\big)Z_{\bb W}^{n_1}	\prod_{A=1}^M Z_{\bb W}(\bar{\bb\sigma}^A  )^{n_2}\Big]^{n_3}=\\&
		=\frac1{\mathcal V}\mE_{\bb \chi}\int d \bb O d{\hat {\bb O}}\exp\Big[-N K(\bb O, \hat{\bb O})+N V(\bb O)-\frac{1-\eta}2 N \sum_\gamma \log \text{det} \big(\bb 1- \frac2{1-\eta }\hat {\bb p}^{\gamma}\big)\\&-\frac{1-\eta}2 N \sum_\gamma \log \text{det} \big(\bb 1- \frac2{1-\eta } {\bb r}^{\gamma}\big)\Big]\sum_{\bb\sigma} \exp\Big(
		\frac1\eta \sum_{\gamma a A i} \hat m ^\gamma _{aA}\sigma_i ^{a\gamma}\bar \sigma_i ^A+\frac1\eta \sum_{\gamma ab i}\hat q^{\gamma }_{ab}\sigma_i ^{a\gamma}\sigma_i^{b\gamma}
		\Big).
	\end{split}
\end{equation}
{We can eliminate the conjugate fields by applying the saddle point method and extremizing the quantity in the exponential} w.r.t. the order parameters, which results in the following condition
$$
\frac{\partial K}{\partial \bb O} = \frac{\partial V}{\partial \bb O},
$$
since the remainder part of the exponential does not depend on the order parameters. Thus, the conjugate fields on the saddle points reads as
\begin{eqnarray}\label{eq:conjs}
	\hat q^\gamma_{ab}&=& \frac{\eta(1-\eta)}{2\beta_2 \epsilon_W}p^\gamma_{ab},\\
	\hat p^\gamma_{ab}&=& \frac{\eta(1-\eta)}{2\beta_2 \epsilon_W}q^\gamma_{ab},\\
	\hat m^\gamma_{aA}&=& \frac{\eta(1-\eta)}{\beta_2\epsilon_W}\sum_{\bar a} n ^\gamma _{a(A\bar a)},\\
	\hat n ^\gamma_{a(A \bar a)} &=& \frac{\eta(1-\eta)}{\beta_2 \epsilon_W}m^\gamma_{aA},\\
 \label{eq:conjs_fin}
	\hat{\bar p}^\gamma_{(A\bar a)(B\bar b)}&=& \frac{\eta(1-\eta)}{2\beta_2 \epsilon_W} q^E_{AB}.
\end{eqnarray}
Upon replacing the expressions \eqref{eq:conjs}-\eqref{eq:conjs_fin} of the conjugate fields in \eqref{eq:r_conj}, one obtains \eqref{eq:r_true}. Notice that we also have
\begin{equation}\label{KOb}
	K(\bb O, \hat{\bb O})= V(\bb O)+\frac{\eta (1-\eta)}{2\beta_2 \epsilon_W}\sum_{\gamma ab} q^\gamma_{ab} p^{\gamma}_{ab}+\frac{\eta(1-\eta)}{\beta_2 \epsilon_W}\sum_{\gamma a A \bar a} m^\gamma_{aA} n^{\gamma}_{a(A\bar a)}.
\end{equation}
and that the final free energy does not depend on ${\bar{\bb p}}$: this is due to the restricted nature of the network we are dealing with (in fact, interactions 
within the hidden layer would break this peculiarity).
After eliminating the conjugate fields and substituting $K(\boldsymbol O,\hat{\boldsymbol O})$ as prescribed by \eqref{KOb}, Eq. \eqref{lastterm} gets

\begin{equation}
\label{eq:final_general}
	\begin{split}
		&\mE _{\bb\chi}	\Big[	\int	d\bb W \exp\big(-\frac{\beta_2N\epsilon_W}{2}\sum_{i\mu} W_{i\mu}^2\big)Z_{\bb W}^{n_1}	\prod_{A=1}^M Z_{\bb W}(\bar{\bb\sigma}^A )^{n_2}\Big]^{n_3}=\\&
		=\frac1{\mathcal V}\int d \bb O \exp\Big[-N\frac{\eta (1-\eta)}{2\beta_2 \epsilon_W}\sum_{\gamma ab} q^\gamma_{ab} p^{\gamma}_{ab}-N\frac{\eta(1-\eta)}{\beta_2 \epsilon_W}\sum_{\gamma a A \bar a} m^\gamma_{aA} n^{\gamma}_{a(A\bar a)}\\&-\frac{1-\eta}2 N \sum_\gamma \log \text{det} \big(\bb 1- \frac\eta{\beta_2 \epsilon_W } {\bb q}^{\gamma}\big)-\frac{1-\eta}2 N \sum_\gamma \log \text{det} \big(\bb 1- \frac2{1-\eta } {\bb r}^{\gamma}\big)\Big]\\&\cdot\mE_{\bb \chi}\sum_{\bb\sigma} \exp\Big(
		\frac{1-\eta}{\beta_2 \epsilon_W} \sum_{\gamma a A \bar ai} n^\gamma_{a(A\bar a)}\sigma_i ^{a\gamma}\bar \sigma_i ^A+\frac{1-\eta}{2\beta_2 \epsilon_W} \sum_{\gamma ab i}p^{\gamma }_{ab}\sigma_i ^{a\gamma}\sigma_i^{b\gamma}
		\Big).
	\end{split}
\end{equation}
\section{Details of computations in the RS regime}\label{app:rs_comp}
In this Appendix, we detail the calculations to reach the expression \eqref{eq:general_replica_Z} appearing in the quenched free energy within the RS ansatz given by Eqs. \eqref{eq:RS_overlaps1}-\eqref{eq:RS_magn2}. First, since there is no explicit dependence on the index $\gamma$, the summation over replicas of third kind is trivial, and results in a factor $n_3$ for the whole quantity in round brackets. Indeed, we have
\begin{equation}
	\begin{split}
		&-N\frac{\eta (1-\eta)}{2\beta_2 \epsilon_W}\sum_{\gamma ab} q^\gamma_{ab} p^{\gamma}_{ab}-N\frac{\eta(1-\eta)}{\beta_2 \epsilon_W}\sum_{\gamma a A \bar a} m^\gamma_{aA} n^{\gamma}_{a(A\bar a)}=
		\\&=-N \frac{\eta(1-\eta)}{2\beta _2 \epsilon_W}n_3n_1 P -N \frac{\eta(1-\eta)}{2\beta _2 \epsilon_W} n_3 n_1(n_1-1)pq-N \frac{\eta(1-\eta)}{\beta _2 \epsilon_W}n_3 n_1 n_2 M m n.
	\end{split}
\end{equation}
Concerning the logarithmic contributions, we first consider
\begin{equation}\label{firstlog}
	\sum_\gamma \log \det \big(\bb 1-\frac{\eta}{\beta_2\epsilon_W}\bb q ^\gamma\big).
\end{equation}
In this case, the eigenvalues of the matrix are
\begin{eqnarray}
	1-\frac\eta{\beta_2 \epsilon_W}(1-q)-n_1 \frac{\eta}{\beta_2\epsilon_W}q \quad & & \quad \text{multiplicity }1 ,\\
	1-\frac\eta{\beta_2 \epsilon_W}(1-q) \quad & & \quad \text{multiplicity } n_1-1.
\end{eqnarray}
Thus, taking the sum over the third family of replica indices, Eq. \eqref{firstlog} becomes
\begin{equation}
\nonumber
	\begin{split}
		\sum_\gamma\log\det\big(\bb 1-\frac{\eta}{\beta_2 \epsilon_W}\bb q^\gamma\big)&= n_3\log\Big(1-\frac{
		\eta}{\beta_2 \epsilon_W}(1-q)-n_1\frac{\eta}{\beta_2 \epsilon_W}q\Big)+n_3(n_1-1)\log \Big(1-\frac\eta{\beta_2 \epsilon_W}(1-q)\Big).
	\end{split}
\end{equation}
The logarithmic contribution 
\begin{equation}
	\sum_\gamma \log\det\big(\bb 1-\frac2{1-\eta}\bb r^\gamma\big)
\end{equation}
 requires more work, as it involves the effective overlap between clamped hidden replicas.
We have to carefully exploit the structure of the matrix $\bb r^\gamma$. First let us recall that, after marginalizing out the conjugate fields of the order parameters on the saddle point, we have
\begin{equation}
	\begin{split}
			r^\gamma_{(A\bar a)(B\bar b)}&=\frac{\eta(1-\eta)}{2\beta_2\epsilon_W}	q_{AB}^E+\frac{\eta^2(1-\eta)}{2\beta_2 \epsilon_W}\sum_{ab} m ^\gamma_{aA} \big(\bb 1 - \frac{\eta}{\beta_2\epsilon_W}{\bb q}^\gamma\big)^{-1}_{ab} m^\gamma_{bB}=\\
   &=
\frac{\eta(1-\eta)}{2\beta_2\epsilon_W}	q_{AB}^E+\frac{\eta^2(1-\eta)}{2\beta_2 \epsilon_W}m^2\sum_{ab} e_a^T \big(\bb 1 - \frac{\eta}{\beta_2\epsilon_W}{\bb q}^\gamma\big)^{-1}_{ab} e_b,
	\end{split}
\end{equation}
where $\bb e$ is the vector of ones. Using \eqref{eq:conjs}, with straightforward computations, one has
\begin{equation}
	\begin{split}
		\sum_{ab} e_a^T \big(\bb 1 - \frac{\eta}{\beta_2\epsilon_W}{\bb q}^\gamma\big)^{-1}_{ab}e_b&=\frac{n_1}{1-\frac{
								\eta}{\beta_2 \epsilon_W}(1-q)-n_1\frac{\eta}{\beta_2 \epsilon_W}q},
	\end{split}
\end{equation}
leading to $r^\gamma_{(A\bar a)(B\bar b)}=	\hat{\bar p}^\gamma_{(A\bar a)(B\bar b)}+\frac{1-\eta}{2}r_0$, where
\begin{equation}
	r_0 = \frac{\eta^2}{\beta_2^2\epsilon_W^2}\frac{n_1m^2}{1-\frac{
			\eta}{\beta_2 \epsilon_W}(1-q)-n_1\frac{\eta}{\beta_2 \epsilon_W}q}.
\end{equation}
Putting all pieces together
\begin{equation}
\begin{split}\label{strm}
	\big(\bb 1-\frac2{1-\eta}\bb r^\gamma\big)_{(A\bar a)(B\bar b)}=\delta_{AB}\delta_{\bar a \bar b}-\frac2{1-\eta}	\hat{\bar p}^\gamma_{(A\bar a)(B\bar b)}-r_0.
\end{split}
\end{equation}
Now, on the saddle point, $\hat{\bar p}^\gamma_{(A\bar a)(B\bar b)}= \eta(1-\eta)q_{AB}^E/2\beta_2\epsilon_W$, and, in the case under consideration, $q_{AA}^E=1$, while $q^E_{AB}=r^2$ plus corrections to the free energy of order $N^{-1/2}$ which can be neglected in the thermodynamic limit. Thus, the relevant matrix has the following structure:
\begin{equation}
	\hat{\bar {\bb p}}^\gamma=
	\begin{pmatrix}
		A & B &B& \dots \\
		B & A & B &\dots\\
		B & B & A & \dots \\
		\vdots & \vdots &\vdots & \ddots
	\end{pmatrix}
\end{equation}
where $A$ and $B$ are $n_2\times n_2 $ matrices of the form
\begin{equation}
	A_{\bar a \bar b} = \frac{\eta(1-\eta)}{2\beta_2\epsilon_W},\quad B_{\bar a \bar b} = \frac{\eta(1-\eta)}{2\beta_2\epsilon_W}r^2.
\end{equation}
It can be shown that the structure of the matrix given in Eq. \eqref{strm} is quite simple; in fact, by setting $n_2=2$ for ease of visualisation, we have
\begin{equation}
	\begin{split}
		&	\bb 1-\frac2{1-\eta}\bb r^\gamma=\\
		=&
		\begin{pmatrix}
			1-\frac{\eta}{\beta_2 \epsilon_W}(1-r^2)& 	-\frac{\eta}{\beta_2 \epsilon_W}(1-r^2)& 	0& 0 &  \\
			-\frac{\eta}{\beta_2 \epsilon_W}(1-r^2)& 	1-\frac{\eta}{\beta_2 \epsilon_W}(1-r^2)&0 &0& \dots \\
			0&0&1-\frac{\eta}{\beta_2 \epsilon_W}(1-r^2)& 	-\frac{\eta}{\beta_2 \epsilon_W}(1-r^2)& \\
			0&0& 	-\frac{\eta}{\beta_2 \epsilon_W}(1-r^2)& 	1-\frac{\eta}{\beta_2 \epsilon_W}(1-r^2)& \\
			 &\vdots  & & & \ddots
		\end{pmatrix}	-\big(\frac\eta{\beta_2 \epsilon_W}r^2+r_0\big).
	\end{split}
\end{equation}
Calling $\bb R$ the block matrix in the last line, we can straightforwardly apply the matrix determinant lemma:
\begin{equation}
	\begin{split}
		\det \big(\bb1-\frac2{1-\eta}\bb r^\gamma\big)&=\det(\bb R)\cdot \big(1-(\frac\eta{\beta_2 \epsilon_W}r^2+r_0)\bb e^T \bb R^{-1}\bb e \big).
	\end{split}
\end{equation}
Because of the block structure of $\bb R$, we trivially have $\det_{Mn_2}({\bb R})= \det_{n_2}({\bb R_0})^M$, where $\bb R_0$ is the $n_2\times n_2$ block on the diagonal. With simple computations, it is easy to show that 
\begin{equation}
	\det (\bb R)= \Big(1-n_2 \frac{\eta}{\beta_2 \epsilon_W}(1-r^2)\Big)^M.
\end{equation}
Further, it can be checked that
\begin{equation}
	\begin{split}
	(\bb R^{-1})_{(A\bar a)(B \bar b)}&=	\frac{1-(n_2-1)\frac{\eta}{\beta_2 \epsilon_W}(1-r^2)}{1-n_2\frac{\eta}{\beta_2 \epsilon_W}(1-r^2)}\delta_{AB}\delta_{\bar a \bar b}+	\frac{\frac{\eta}{\beta_2 \epsilon_W}(1-r^2)}{1-n_2\frac{\eta}{\beta_2 \epsilon_W}(1-r^2)}\delta_{AB}(1-\delta_{\bar a \bar b}),
	\end{split}
\end{equation}
leading to $\bb e^T \bb R^{-1}\bb e=\frac{Mn_2}{1-n_2 \frac{\eta}{\beta_2 \epsilon_W}(1-r^2)}$. Then, putting everything together, we have
\begin{equation}
	\begin{split}
			&\log\det \big(\bb1-\frac2{1-\eta}\bb r^\gamma\big)=\\=&\log\Big[1-\frac\eta{\beta_2 \epsilon_W}\frac{Mn_2 r^2}{1-n_2 \frac{\eta}{\beta_2 \epsilon_W}(1-r^2)}-\frac{Mn_2}{1-n_2 \frac{\eta}{\beta_2\epsilon_W}(1-r^2)}\frac{\eta^2 }{\beta_2^2 \epsilon_W^2}\frac{n_1 m^2}{1-\frac{\eta}{\beta_2 \epsilon_W}(1-q)-n_1 \frac{\eta}{\beta_2 \epsilon_W}q}\Big],
	\end{split}
\end{equation}
where in the last line we neglected unessential constant terms which do not cause issues for a proper extremization of the quenched free energy in the $M\to\infty$ limit. Summation over $\gamma$ is again trivial.
\par\medskip
Finally, the quantity involving the free replica of the visible layer can be treated with standard techniques in replica trick computations. Exploiting the RS ansatz \eqref{eq:RS_overlaps1}-\eqref{eq:RS_magn2}, it can be checked that, up to $\mathcal O(n_3^2)$ {corrections}, the quantity in the last line of Eq. \eqref{eq:final_general} reduces to
\begin{equation*}
{	\begin{split}
	&\mE_{\bb\chi}\sum_{\bb\sigma} \exp\Big[
	\frac{1-\eta}{\beta_2 \epsilon_W} \sum_{\gamma a A \bar ai} n^\gamma_{a(A\bar a)}\sigma_i ^{a\gamma}\bar \sigma_i ^A+\frac{1-\eta}{2\beta_2 \epsilon_W} \sum_{\gamma ab i}p^{\gamma }_{ab}\sigma_i ^{a\gamma}\sigma_i^{b\gamma}\Big]=\exp\Big\{\frac{\eta(1-\eta)}{2\beta_2\epsilon_W} N n_3 n_1 (P-p)\\&+n_3n_1\eta N \log 2+n_3\eta N \int d\mu(z)\log		\int d\mu(\tau)\cosh^{n_1}	\big[\frac{1-\eta}{\beta_2\epsilon_W}Mn_2r(1+\sqrt \rho z)+\sqrt{\frac{1-\eta}{\beta_2\epsilon_W}p}\, \tau \big]\Big\},
\end{split}}
\end{equation*}
where, since $\sigma_i ^E \sim\xi_i r\big(1+	\sqrt{\rho}z\big)$ in the $M\to\infty$ limit we straightforwardly replace  $\mathbb{E}_{\boldsymbol\chi}(\cdot)\to\int d\mu(z)(\cdot)$. 

\section{General replica-symmetric Ansatz}\label{app:general_RS}
In this Appendix, we inspect a more general ansatz, in which the overlap order parameters display a RS structure, while we retain the magnetizations depending on the example index. This is translated in mathematical terms as
\begin{eqnarray}
\label{eq:new-ansatz1}
	m^\gamma_{aA}&=& m^A,\\
	n^{\gamma}_{a(A\bar a)}&=& n^A,\\
	q_{ab}^\gamma&=& (1-q)\delta_{ab}+q,\\
 \label{eq:new-ansatz2}
	p_{ab}^\gamma &=& (P-p)\delta_{ab}+p.
\end{eqnarray}
With similar (and cumbersome) computations as in the usual RS ansatz used in the main text, the argument of the exponential in the Laplace-integral form of the quenched free energy is straightforwardly generalized as
\begin{equation}
		\begin{split}
			\frac{f_M (\bb O)}{\eta}&=-\frac{\alpha\beta_\epsilon}{2 }(\beta_2+1)pq+{\alpha\beta_\epsilon} \beta_2\frac1M \bb  m \cdot  \bb n+\frac{\alpha\beta_\epsilon}{2 } p+\frac{\alpha(\beta_2+1)}{2} \log \Big[1-\frac{\beta_\epsilon}{\beta_2 }(1-q)\Big]\\
			&-\frac{\alpha}{2}
			\log		\Big\{\Big(1-{}{ }\frac{ \beta_\epsilon  r^2}{1-{ \beta_\epsilon }{ } r^2 \rho}\Big)\Big[1-\frac{
				\beta_\epsilon}{\beta_2 }(1-q)+{\beta_\epsilon}{}q\Big]+{\beta_\epsilon^2}{} { \frac1M\bb m ^T\tilde {\bb g }\bb m}\Big\}
			 \\&+\int d\mu(z)\log\int d\mu(\tau) \cosh^{-\beta_2}\big({\alpha\beta_\epsilon}{ } r \frac1M \sum_A n^A +{\alpha\beta_\epsilon}{} r \sqrt {\rho }   \frac{\lVert \bb n \lVert }{\sqrt M}z+\sqrt{\frac{\alpha\beta_\epsilon}{\beta_2}p}\tau\big),
		\end{split}
\end{equation}
%%%
where we already performed the analytical continuation on $n_1$ and $n_2$ and took their relevant limits (while keeping explicit $r$, $\rho$ and $M$). The matrix
$$
\tilde g_{AB}=\delta_{AB} \frac{1-{ \beta_\epsilon }{}r^2(\rho+ 1)}{(1-{ \beta_\epsilon }{ }r^2\rho)^2}+\frac{ \beta_\epsilon }{M }\frac{r^2}{(1-{ \beta_\epsilon }r^2\rho)^2}
$$
is proportional to the inverse of $(\bb 1-\frac{\beta_\epsilon}{M}\bb q^E)$, and plays the role of metric tensor in the magnetization space. As a consequence, its spectral properties are crucial to understand the scenario. In particular, $\tilde{\bb g}$ is a linear combination of the identity matrix of dimension $M$ and the matrix of ones; as a consequence, there are only two possible eigenvalues:
\begin{eqnarray}
\lambda_1&=&	\frac{1}{1-r^2 \beta_\epsilon \rho}\quad \text{multiplicity }1,\\
\lambda_0&=&	\frac{1-r^2 \beta_\epsilon (1+\rho)}{(1-r^2 \beta_\epsilon \rho)^2}\quad \text{multiplicity }M-1.
\end{eqnarray}
The associated eigenspaces are $\textrm{Span}\{\bb v^{(1)}=(1,1,\dots,1)\}$ and $\textrm{Span}\{\bb v^{(2)},\dots,\bb v^{(M)}\}$ with $\sum_A v^{(B)}_A=0$ for $B\ge2$, respectively. We further stress that, in the scenario we deal with (namely $r=(1+\rho_0)^{-1/2}$ and $\rho =0$, so that $M\to\infty$), the lowest eigenvalue is $\lambda_0=1-\beta_\epsilon/(1+\rho_0)$, while $\lambda_1 =1$. {As explained in the main text, the condition $\lambda_0=0$ corresponds to the regularization threshold} 
(note that, the matrix $\tilde{\bb g}$ becomes singular there), so that $\lambda_0<0$ naturally identifies the low-regularization regime.\par\medskip
Extremizing the quenched free-energy, one easily gets the following set of self-consistency equations:
\begin{eqnarray}
	m^A &=& r\tilde m+\alpha r^2 \rho \tilde n (1-\beta_\epsilon r^2\rho)\big(1-\beta_2(q-\tilde q)-q\big)\sum_B \tilde g_{AB} m^B,\\
 n^A  &=& \frac{\tilde n (1-\beta_\epsilon r^2 \rho)}{\beta_\epsilon}\sum_B \tilde g_{AB}m^B,\\
	 \tilde n&=&\frac{T_2}{\frac{  (1-\beta_\epsilon r^2 \rho)}{M}\bb m^T \tilde{\bb g} \bb m-\Big[{ r^2(1+\rho)}-\frac1{\beta_\epsilon}\Big]\Big[\frac1{\beta_\epsilon}-T_2(1-q)+ q \Big]},\\
	 \tilde p &=&\frac{T_2^2}{\frac1{\beta_\epsilon}-T_2(1-q)}+\tilde n T_2 (r^2 (1+\rho)-\frac1{\beta_\epsilon})			,\\
	 \tilde m &=& I_1=\int d\mu(z)\frac{\int d\mu(\tau) \cosh^{-\beta _2}(g(z,\tau))\tanh(g(z,\tau))}{\int d\mu(\tau)\cosh^{-\beta_2}(g(z,\tau))},\\
	 q &=& I_2^{(1)}=\int d\mu(z)\frac{\int d\mu(\tau) \cosh^{-\beta _2}(g(z,\tau))\tanh^2(g(z,\tau))}{\int d\mu(\tau) \cosh^{-\beta_2}(g(z,\tau))},\\
	 \tilde q &=& I_2^{(2)}=\int d\mu(z) \Big(\frac{\int d\mu(\tau) \cosh^{-\beta _2}(g(z,\tau))\tanh(g(z,\tau))}{\int d\mu(\tau)\cosh^{-\beta_2}(g(z,\tau))}\Big)^2,
\end{eqnarray}
with $g(z,\tau)=\alpha \beta_\epsilon r \frac1M \sum_A n^A +\alpha\beta_\epsilon r \sqrt \rho\frac{\lVert \bb n\lVert }{\sqrt M}z +\sqrt{\alpha \tilde p}\tau$. These intricate equations constitute the generalization of \eqref{eq:n_final_0}-\eqref{eq:q_final_0} for the new ansatz. Let us now focus again on the equation for the magnetization w.r.t. the examples, and consider it on the saddle point (we denote the solution with the * symbol for the order parameters), in matrix form we have:
$$
(\bb 1-f_{q,\tilde q ,\tilde n} \tilde {\bb g}{ } )\bb {m}^* = r\tilde m^* \bb v^{(1)} 
$$
where $f_{q,\tilde q,\tilde n}=\alpha r^2 \rho \tilde n^* (1-\beta_\epsilon r^2\rho)\big(1-\beta_2(q^*-\tilde q^*)-q^*\big)$ and we recall that $\bb v^{(1)}=(1,1,\dots,1)$ is the top eigenvector of $\tilde{\bb g}$. This is clearly a linear system in $\bb m^*$, and:
\begin{enumerate}
	 \item provided that the matrix $\bb 1 -f_{q,\tilde q,\tilde n }\tilde{\bb g}$ is non-singular, the solution is unique. Such solution is precisely in the form $\bb m^*= m^* \bb v^{(1)}$, so that the linear system reduces to the condition
	$$
	[1-\alpha r^2 \rho \tilde n^* \big(1-\beta_2(q^*-\tilde q^*)-q^*\big)] m^*=r \tilde m^*,
	$$
	which is the same as the special RS ansatz treated in the main text. Conversely, if $\tilde m^*=0$, the only possible solution is ${m^*}^A=0$ since the system is homogeneous.
	\item 
	
	If the matrix $\bb1 -f _{q,\tilde q,\tilde n}\tilde{\bb g}$ is singular and $\tilde m^*\neq 0$, the system has no solution, since the complete matrix $(\bb1 -f _{q,\tilde q,\tilde n}\tilde{\bb g}\vert r \tilde m\bb v^{(1)})$ is full-rank (the determinant of $\bb1 -f _{q,\tilde q,\tilde n}\tilde{\bb g}$ with a column replaced with $r \tilde m^*$ has determinant $\propto \tilde m^*$), and thus the linear system is incompatible.\\
	
	\item Finally, consider the case of $\bb1 -f _{q,\tilde q,\tilde n}\tilde{\bb g}$ being singular and $\tilde m^*=0$. In this case, $\bb m^*$ belongs to the kernel of $\bb1 -f _{q,\tilde q,\tilde n}\tilde{\bb g}$. The matrix $\bb1 -f _{q,\tilde q,\tilde n}\tilde{\bb g}$ is singular in two cases:
	
	\begin{itemize}
		\item $f_{q,\tilde q,\tilde n}\lambda^1=1$, with $\lambda^1=(1-\beta_\epsilon r^2\rho)^{-1}$ (mult. 1). In this case, the matrix $\bb1 -f _{q,\tilde q,\tilde n}\tilde{\bb g}$ has rank $M-1$, and  the kernel is the subspace in $\mathbb R^M$ of vectors collinear with $\bb v^{(1)}$, thus $\bb m^*=m^*\bb v^{(1)}$, reducing to the special RS ansatz; notice that, in our regime $r^2=(1+\rho_0)^{-1}$ and $\rho=0$, this means $f_{q,\tilde q,\tilde n}=1$, which is equivalent to say that $\tilde n \sim (r^2\rho)^{-1} \propto M^{-1} \to 0$ in the large dataset limit. In terms of the order parameters, this would lead the denominator in the self-consistency equation for $\tilde n$ to be zero, which means that the argument of the second logarithm in the free energy should be zero: this is a singular solution of the saddle point equations, and lies outside the existence conditions of a proper extremal point, thus it should be excluded.
		\item $f_{q,\tilde q,\tilde n}\lambda^0=1$, with $\lambda^0=\frac{1-\beta_\epsilon r^2 (\rho+1)}{(1-\beta_\epsilon r^2 \rho)^2}$ (mult. $M-1$). In this case, the matrix $\bb1 -f _{q,\tilde q,\tilde n}\tilde{\bb g}$ has rank 1, and the kernel is the span of all linear combinations of eigenvectors $\bb v^{(A\ge 2)}$, thus the sum of entries of vectors in this subspace sum to zero, that is $\frac1M \sum_A {m^*}^A=0$. However, this solution is not admissible too, since it contradicts the conditions of proper existence of the saddle point. Indeed, let us focus on the inequalities:
		\begin{eqnarray}
		1-\frac{\beta_\epsilon}{\beta_2}(1-q)&>&0,\label{eq:app_cond1}\\
		\Big(1-{}{ }\frac{ \beta_\epsilon  r^2}{1-{ \beta_\epsilon }{ } r^2 \rho}\Big)\Big[1-\frac{
			\beta_\epsilon}{\beta_2 }(1-q)+{\beta_\epsilon}{}q\Big]+{\beta_\epsilon^2}{} { \frac1M\bb m ^T\tilde {\bb g }\bb m}&>&0,
		\end{eqnarray}
	coming from the two logarithms in the free energy. The last inequality can be rewritten as
	$$
	\frac{\lambda_0}{\lambda_1}\delta_q+{\beta_\epsilon^2}{} { \frac1M\bb m ^T\tilde {\bb g }\bb m}>0
	$$
	with $\delta_q= 1-\frac{
		\beta_\epsilon}{\beta_2 }(1-q)+{\beta_\epsilon}{}q$. Since $q>0$, it is straightforward to see that $\delta_q>0$ by virtue of \eqref{eq:app_cond1}. Now, if $\bb m$ lies in the eigenspace of $\lambda_0$, we have
	$$
	\frac{\lambda_0}{\lambda_1}\delta_q +\beta_\epsilon^2 \frac1M \lambda_0 \lVert \bb m \lVert ^2>0.
	$$
	Since above the regularization threshold $\lambda_0<0$, we have
$$	
\frac{1}{\lambda_1}\delta_q +\beta_\epsilon^2 \frac1M  \lVert \bb m \lVert ^2<0 \Rightarrow \delta_q <- \frac{\lambda_1 \beta_\epsilon^2}{M }\lVert \bb m\lVert ^2\le0,
$$
which is incompatible with \eqref{eq:app_cond1}. Thus, {the current case does not provide} a proper extremal point of the free energy.

\end{itemize}
\end{enumerate}
The above-discussion highlights that, as long as we are interested in the sampling regime, the general RS ansatz \eqref{eq:new-ansatz1}-\eqref{eq:new-ansatz2} exactly reduces to the one addressed in the main text.

\section{Critical transitions below the regularization threshold}\label{app:critical}
{In this Appendix, we will give an analytical argument for the existence of critical transitions below the high regularization regime ($\beta_\epsilon<1+\rho_0$) starting from the self-consistency equations \eqref{eq:n_final_0}-\eqref{eq:q_final_0}. We stress that, following this route does not ensure the proper extremization conditions of the quenched free energy, as self-consistency equations are only derived setting to zero the gradient of the free-energy w.r.t. the order parameters, thus without checking that the corresponding saddle-point is well-defined. Thus, the results here reported should be read in combination with the discussion in Subsec. \ref{subsec:theo_cons}. We start our discussion by analyzing the spin-glass solution (i.e. setting $m=0$ in the self-consistency equations). In this case, the solution does not depend on $\rho_0$. A proper expansion of the order parameter $\tilde p$ in powers of $q$ reads as}
$$
\tilde p =\frac{\sqrt{\alpha } \beta_ \epsilon T_2\tau  }{\beta _\epsilon  T_2-1} \sqrt{q}+\frac{\sqrt{\alpha } \beta _\epsilon ^2 \tau  T_2 (2 T_2+1)}{2 (\beta_ \epsilon  T_2-1)^2} q^{3/2}+\frac{\sqrt{\alpha } \beta_ \epsilon ^3 \tau  T_2 (8 T_2 (T_2+1)+3)}{8 (\beta_ \epsilon  T_2-1)^3}q^{5/2} +\mathcal O\left(q^{7/2}\right).
$$
By using this expansion, we can recast the self-consistency equation for $q$ as
\begin{equation*}
	\begin{split}
		q &= I_2 =\frac{\alpha  \beta _\epsilon ^2 T_2^2}{(\beta_ \epsilon  T_2-1)^2}q -\frac{\alpha  \beta_ \epsilon ^3  T_2^2 (2 T_2+1) ((\alpha -1) \beta_ \epsilon  T_2+1)}{(\beta _\epsilon  T_2-1)^4}q^2+ \mathcal O\left(q^{5/2}\right).
	\end{split}
\end{equation*}
Limiting our analysis around the critical point, we can neglect higher-order terms and solve the equations up to order $q^2$. Clearly, we have two branches of solutions, the trivial one $q=0$ and
\begin{equation}
	q=\frac{(1-\beta _\epsilon  T_2)^2 [\beta_ \epsilon  T_2 ((\alpha -1) \beta_\epsilon  T_2+2)-1]}{\alpha  \beta_ \epsilon ^3 T_2^2 (2 T_2+1) ((\alpha -1) \beta_ \epsilon  T_2+1)}\approx 2\Big(1-\frac{2}{\beta_\epsilon(\sqrt{\alpha }+1)+2}\Big) (T_2-T_c^{SG}),
\end{equation}
where\footnote{Notice that there is another branch $T^-_c= \beta_\epsilon^{-1}(1-\sqrt \alpha)^{-1}$, which is unphysical (i.e., it is above $T_c^{SG}$ or it is negative) and can be excluded.}
$$
T_c ^{SG}(\alpha,\beta_\epsilon)= T_c^+=\beta_\epsilon^{-1}(1+\sqrt \alpha)^{-1}.
$$
In particular, we see that, at $\alpha=0$ we have $T_c ^{SG}=\beta_\epsilon^{-1}$, while for $\alpha\to\infty$ we have $T_c^{SG}=0$.
Also, notice that we can invert the previous equation in order to get the critical parameter $\alpha_c(T_2)$, as
$$
\alpha_c ^{SG}(T_2,\beta_\epsilon)=\frac{(1-\beta_ \epsilon  T_2)^2}{\beta _\epsilon ^2 T_2^2}, \quad \text{ for } T_2\in [0,\beta_\epsilon^{-1}].
$$
Around the critical point, the spin-glass solution is approximated as
\begin{equation}\label{eq:sg_crit}
	q\approx \frac{T_2 ^3 \beta_\epsilon^2}{(1+2T_2)(1-T_2\beta_\epsilon)}(\alpha-\alpha_c ^{SG}).
\end{equation}
The spin-glass solution is non-trivial for $\alpha\ge \alpha_c ^{SG}(T_2{,\beta_{\epsilon}})$, otherwise it coincides with the ergodic one. Since $\alpha_c^{SG}$ is defined for $T_2 \in [0,\beta_\epsilon^{-1}]$ (i.e. $\beta_\epsilon<\beta_2$), for $T_2 >\beta_\epsilon^{-1}$ critical transitions are not expected.
\begin{figure}
	\centering
	\begin{minipage}{0.49\textwidth}
		\centering
		\includegraphics[width=\textwidth]{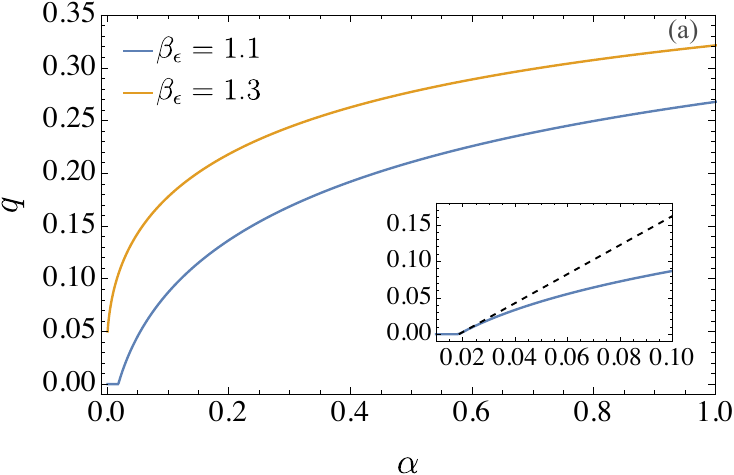}
	\end{minipage}
	\begin{minipage}{0.49\textwidth}
	\centering
	\includegraphics[width=\textwidth]{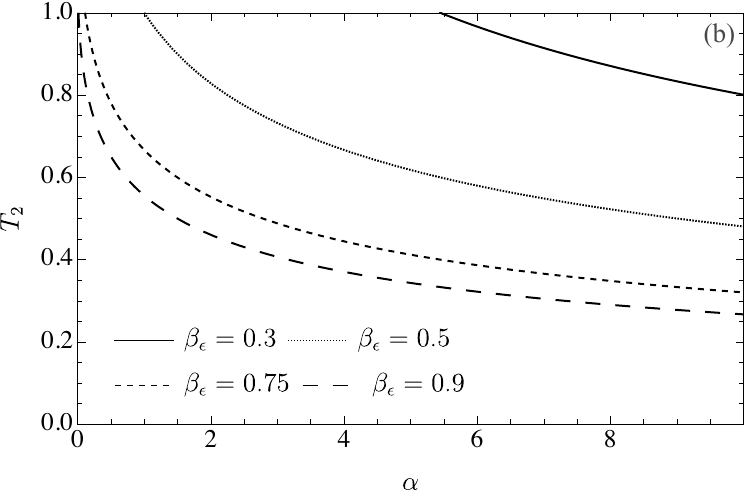}
\end{minipage}\\
	\begin{minipage}{0.49\textwidth}
	\centering
	\includegraphics[width=\textwidth]{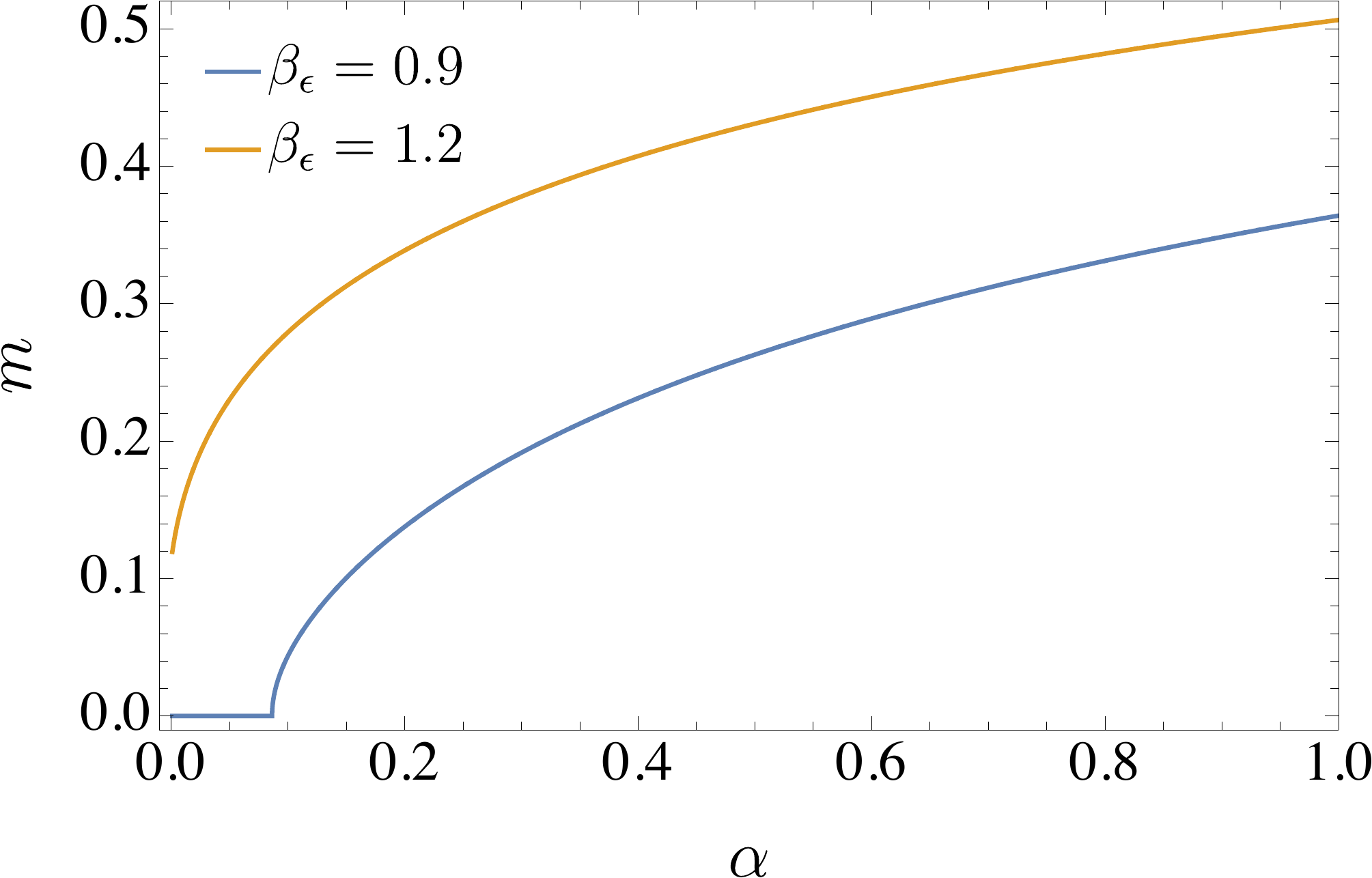}
\end{minipage}
\begin{minipage}{0.49\textwidth}
	\centering
	\includegraphics[width=\textwidth]{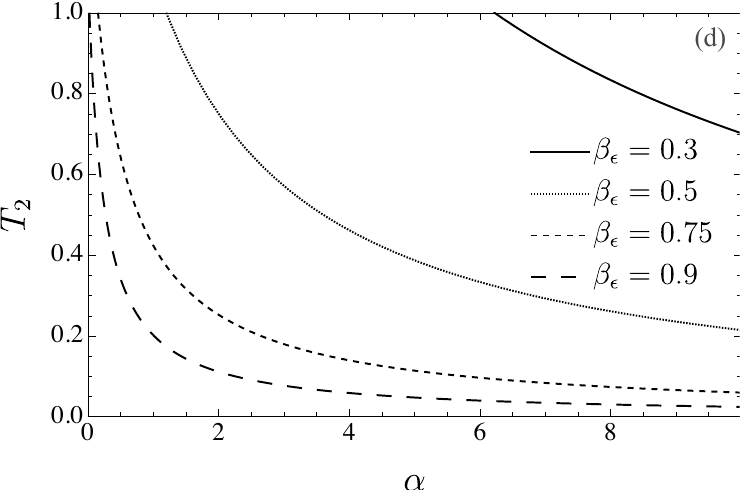}
\end{minipage}
\caption{{\bfseries Critical transitions for spin-glass and retrieval solutions.} Panel a) The plot shows the spin-glass solution at $T_2=0.8$ as a function of $\alpha$ for $\beta_\epsilon=1.1$ (blue curve) and $\beta_\epsilon=1.3$ (orange curve). In the first case, we have $\beta_\epsilon^{-1}\approx 0.91$, so $T_2 <\beta_\epsilon^{-1}$, and the spin-glass solution exhibits a critical transition; in the second case, $\beta_\epsilon^{-1}\approx 0.78$, so $T_2 >\beta_\epsilon^{-1}$, and {the overlap for} the spin-glass solution {never vanishes}. This is in agreement with the theoretical analysis. In the inset, we reported a detail of the solution at $\beta_\epsilon=1.1$ and compare it to the analytic estimation of the overlap order parameter around the critical point, as given by Eq.~\eqref{eq:sg_crit}. Around the critical point, the two curves are in perfect agreement. Panel b) The plot shows the critical curve at which the model exhibits a non-zero overlap for the spin-glass solution for various values of the regularization parameter $\beta_\epsilon$, in a region for which is ensured that $T_2 <\beta_\epsilon^{-1}$. Panel c) The plots show two examples of the magnetization as a function of $\alpha$ for $T_2=0.8$ and $\rho_0=0.1$ for $\beta_\epsilon=0.9$ and $\beta_\epsilon=1.2$. In the former case, we have a critical transition for $m$, while for the latter the solution is always non-zero. Panel d) The plot shows the critical curves for the transition of the retrieval solution at $\rho_0=0.1$ under the assumption that it experience a critical behavior before the spin-glass solution. This is in perfect agreement with our numerical results. Further details about the retrieval solution are given in the main text and in Fig. \ref{fig:whole_crit}.} 
\end{figure}
\par\medskip
Let us now focus on the retrieval solution and expand the self-consistency equations around $m=0$. When $m\to 0$, the retrieval and the spin-glass solutions will merge together. Thus, one has to check the emergence of a critical transition in $m$ by expanding around the spin-glass solution. However, we choose another route, that is: we assume that, around $m=0$, the spin-glass solution is absent (i.e., $q^{SG}=0$), and check {\it a posteriori} if (and where) this assumption is satisfied. If this is not the case, we will find a critical behavior for $m$ only, while the overlap in the retrieval solution for $q$ will merge with the spin-glass one. Expanding around $\tilde m=0$, the argument of the hyperbolic functions can be approximated as
\begin{equation*}
	g(\tau) = g_0\tau + g_1 \tilde m +g_2 \tau \tilde m^2+g_3 \tilde m^3+\dots
\end{equation*}
where $g_n = g_n(q; \alpha,\beta_\epsilon, T_2,\rho_0) $. Even terms in the expansion linearly depends on $\tau$, while odd ones are independent (this is due to the fact that both $\tilde n$ and $\tilde p$ are expanded in even powers of $\tilde m$, but the former enters in $g$ through the factor $\tilde n m$). In particular, we find
\begin{eqnarray}
	g_0 &=& \frac{\sqrt{\alpha q } \beta _\epsilon   {T_2}}{\sqrt{[\beta_\epsilon {T_2} (q-1) +1 ][\beta_ \epsilon  ( {T_2}(q-1)+q)+1]}},\\
	g_1 &=&\frac{\alpha  \beta _\epsilon ^2 {T_2}}{( 1+\rho _0-\beta_ \epsilon) [\beta_ \epsilon  (T_2(q-1) +q)+1]},\\
	g_2 &=&\frac{\sqrt{\alpha } \beta _\epsilon ^2 {T_2} \sqrt{\beta_ \epsilon T_2 (q-1)+1}}{2\sqrt{q} (1+\rho_0 - \beta_ \epsilon) [\beta _\epsilon  (T_2(q-1) +q)+1]^{3/2}}.
\end{eqnarray}
When expanding the integrals {$I_1$ and $I_2$} around $\tilde m=0$, we found non-trivial terms to evaluate such as
$$
\mu_4 = \frac{\mE_\tau\cosh^{-\beta_2}(g_0 \tau)\tanh^4(g_0\tau)}{\mE_\tau \cosh^{-\beta_2}(g_0 \tau)}.
$$
In general, they should be evaluated for each choice of the tunable parameters $\alpha,\beta_\epsilon$ and $T_2$. However, under the assumption that the retrieval solution undergoes a critical transition before the spin-glass one (i.e., $q^{SG}=0$ around the critical point of the retrieval solution), and using $\tanh^4(z)\le \tanh^2(z)$ for all $z\in \mathbb R$, we have
$$
0\le \mu_4 = \frac{\mE_\tau\cosh^{-\beta_2}(g_0 \tau)\tanh^4(g_0\tau)}{\mE_\tau \cosh^{-\beta_2}(g_0 \tau)}\le \frac{\mE_\tau\cosh^{-\beta_2}(g_0 \tau)\tanh^2(g_0\tau)}{\mE_\tau \cosh^{-\beta_2}(g_0 \tau)}= q^{SG}=0,
$$
thus $\mu_4=0$. With these tricks, we can evaluate
\begin{eqnarray}
I_1 &=&{g_1} \tilde m+ \big({g_3}-\frac13 {g_1^3}\big) \tilde m^3 +\mathcal O(\tilde m^5),\\
I_2 &=&g_1^2 \tilde m^2 + \big(-\frac{2 g_1^4}{3}+2 {g_1} {g_3}+g_2^2\big) \tilde m^4+O(\tilde m^6).
\end{eqnarray}
Using these expansions, it is possible to see that, up to the third order, and recovering the example magnetization $m$, we get the equation
\begin{equation*}
{	\begin{split}
	&m \Big(\frac{\alpha  \beta_ \epsilon ^2 T_2}{(1 +\rho_0-\beta_ \epsilon) (\beta_ \epsilon  T_2-1)}+1\Big)-m^3\frac{\alpha  \beta_ \epsilon ^4 (\rho_0+1) T_2}{3 (\beta_ \epsilon -\rho_0-1)^3 (\beta_ \epsilon  T_2-1)^4}\cdot\\
	 &\cdot\Big({3 (\rho_0+1)+\beta_ \epsilon ^3 T_2^2 \left(\alpha ^2 (2 T_2+3)-3\right)+\beta_ \epsilon ^2 T_2 \left(T_2 \left(\alpha ^2+3 \rho_0+3\right)+6\right)-3 \beta_ \epsilon  (2 (\rho_0+1) T_2+1)}\Big)=0.
	\end{split}}
\end{equation*}
The previous equation has two kinds of solution, the trivial one $m=0$ and $m=\pm \vert m_c\vert$. It can be checked that $m_c$ does vanish when $T_2$ equals
$$
T_c ^R (\alpha,\beta_\epsilon)=\frac{1+\rho_0-\beta _\epsilon }{\beta _\epsilon  \left(1+\rho_0-\beta_\epsilon +\alpha  \beta _\epsilon \right)}.
$$
The critical transition therefore occurs at $T_c ^R$, which plays the role of critical temperature of ergodicity-breaking phase transition (due to our assumption $\alpha_c ^R< \alpha_c ^{SG}$). Again, we see that $T_c ^R\in [0,\beta_\epsilon^{-1}]$. The expansion of $m_c$ around the critical transition reads as
\begin{equation*}
	m  \approx \pm \frac{\sqrt{ 3T_2 }\beta_\epsilon }{\sqrt{1+\rho_0} \sqrt{2T_2 \beta_\epsilon(1+\rho_0-\beta_\epsilon)+1+\rho_0 +\beta_\epsilon(2+3\rho_0)}  }  \sqrt{\alpha- \alpha_c^R},
\end{equation*}
with
\begin{equation}
	\alpha_c ^R (T_2,\beta_\epsilon)= \frac{(1-T_2\beta_\epsilon)(1+\rho_0-\beta_\epsilon)}{T_2 \beta_\epsilon^2}.
\end{equation}
Notice also that, in order for the critical transition to exist, we should have
\begin{equation}\label{eq:condT2}
	T_2 > \frac{1+\rho_0 +\beta_\epsilon(2+3\rho_0)}{2\beta_\epsilon(\beta_\epsilon-1-\rho_0)}.
\end{equation}
\begin{figure}[h!]
	\centering
	\includegraphics[width=0.95\textwidth]{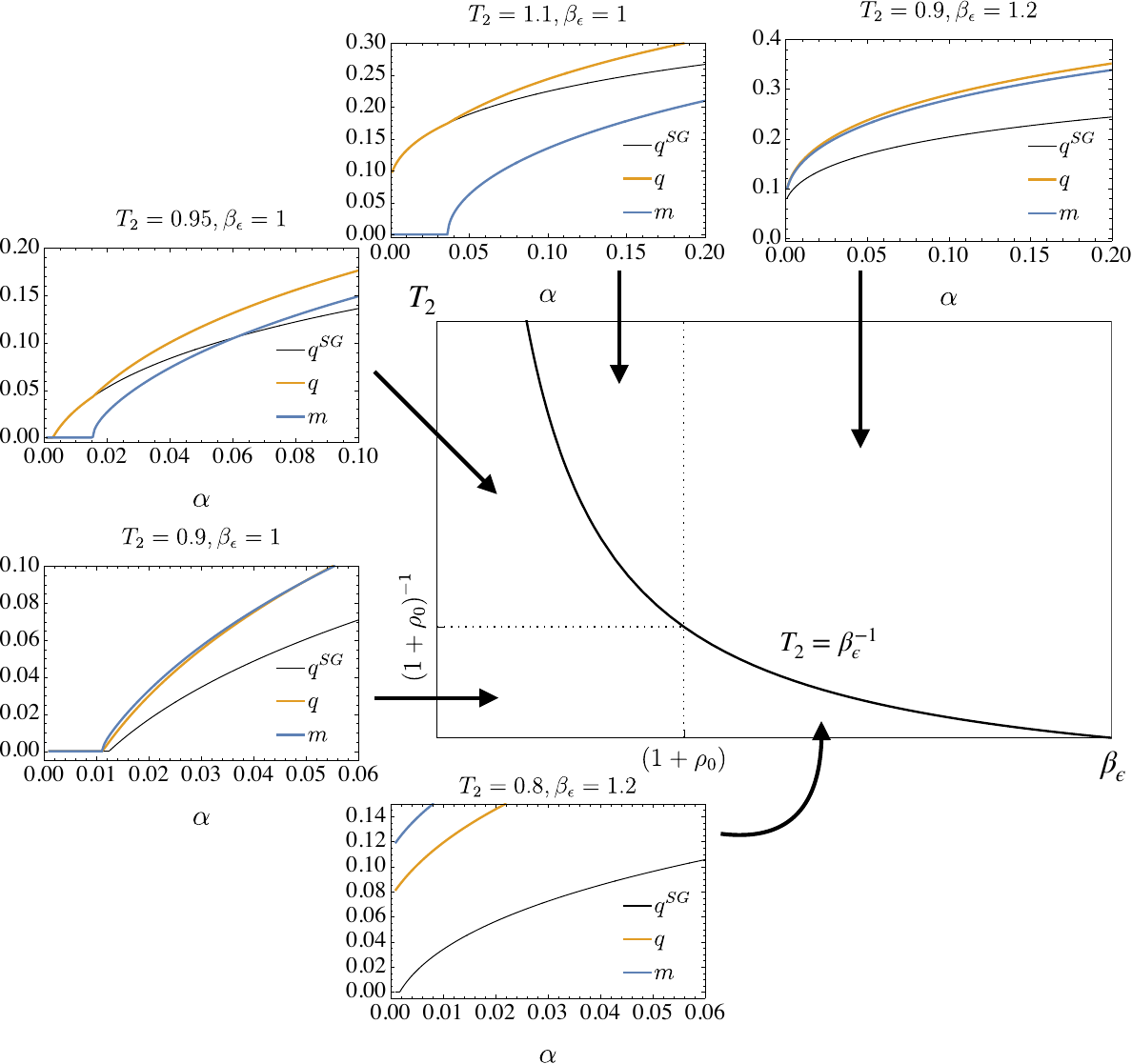}
	\caption{{\bfseries Schematical representation of critical transitions for the BM.} The figure shows the different regimes in terms of the parameters $T_2$ and $\beta_\epsilon$, giving an example of solution both for the retrieval and the spin-glass case (again, no arguments about the free energy is given, we only solve the self-consistency equations). The solutions are realized for $\rho_0=0.1$.}\label{fig:whole_crit}
\end{figure}
If $\beta_\epsilon<1+\rho_0$ (i.e. below the regularization threshold), the condition is always satisfied, and we expect that critical transitions for the retrieval solution always hold in the range $T_2\in [0,\beta_\epsilon^{-1}]$. In the opposite case, above the regularization threshold ($\beta_\epsilon>1+\rho_0$), the r.h.s. in the previous inequality is always greater than $\beta_\epsilon^{-1}$, thus there are no temperatures $T_2$ leading to critical transitions. Finally, the condition $\alpha_c^R\le\alpha_c ^{SG}$ for the retrieval solution to experience a critical transition before the spin-glass one leads to $T_2 < (1+\rho_0)^{-1}.$

To summarize, we have different {scenarios for the self-consistency equations}:
\begin{itemize}
	\item {\bfseries Below the regularization threshold:} $\beta_\epsilon<1+\rho_0$. If $T_2 <(1+\rho_0)^{-1}$, then both the retrieval and the spin-glass solution will experience a critical transition, and in particular the former has a lower critical capacity. The expansion of $m$ and $q$ around $0$ is therefore applicable. If $(1+\rho_0)^{-1}<T_2 < \beta_\epsilon^{-1}$, again both the solutions exhibit critical behavior, but the spin-glass solution has a lower critical capacity: the retrieval solution would therefore be expanded around this point. Finally, if $T_2 >\beta_\epsilon ^{-1}$, the spin-glass solution would not exhibit a critical behavior, but the magnetization for the retrieval solution can still exhibit it (the condition $T_2 <\beta_\epsilon^{-1}$ was only possible under the assumption $\alpha_c ^R< \alpha_c ^{SG}$).
	\item  {\bfseries Above the regularization threshold:} $\beta_\epsilon>1+\rho_0$. The condition $T_2<\beta_\epsilon^{-1}$ implies $T_2<(1+\rho_0)^{-1}$, thus the retrieval solution would experience a transition before the spin-glass one; however, the condition \eqref{eq:condT2} is violated, thus both $m$ and $q$ would be non vanishing. Conversely, the spin-glass solution is critical. On the other hand, if $T_2>\beta_\epsilon^{-1}$, also the overlap for the spin-glass solution is non-vanishing (with the retrieval one being always located at some $\alpha<\alpha_c^{SG}$); this means that, here, critical transitions are forbidden.
\end{itemize}
To conclude this Appendix, we stress that, when training a BM, one would of course choose the tunable parameters in order to avoid critical phase transitions: indeed, we would like to have a machine starting to well-generalize {even for low relative size between the layers}. Critical transitions of course do not satisfy this requirement, since reaching higher and higher values of $m$ would require very large values of the parametrization loads $\alpha$.

\section{Computing training performances}\label{app:train_perfs}
In our numerical simulations, we performed an extensive number of training procedures on networks with different size and values of the hyper-parameters. Thus, the computation of the KL divergence -- even {applying approximations to estimate} the partition function of the model -- is unfeasible. To this aim, we followed a pseudo-loglikelihood approach based on conditional probabilities. Since we want to estimate the target distribution $Q(\bb\sigma)$ with the model marginal distribution $P_{\bb W}(\bb\sigma)$, we consider the factorization
$$
P_{\bb W} (\bb \sigma =\bar { \bb \sigma}^A)=  P_{\bb W} (\sigma_k = \bar\sigma _k ^A \vert\bar{ \bb \sigma}_{\backslash k}^A) P_{\bb W} (\bar{\bb\sigma}_{\backslash k}^A),
$$
where $\bar{\bb\sigma}_{\backslash k}^A= (\bar \sigma_ 1 ^A,\dots, \bar \sigma_{k-1}^A,\bar \sigma_{k+1}^A,\dots, \bar \sigma_{N_\sigma}^A)$ {represents the configuration of the visible units where all the entries, but the $k$-th one, are fixed on the generic sample $A$ and $P_{\bb W} (\bar{\bb\sigma}_{\backslash k}^A)$ is the related marginalized distribution}:
\begin{equation*}
	\begin{split}
		P_{\bb W} (\bar{\bb\sigma}_{\backslash k}^A)&= \sum_{\sigma_k =\pm1} P_{\bb W}(\bb\sigma)\Big\vert_{\bb\sigma _{\backslash k}= \bar{\bb\sigma}_{\backslash k}^A}= Z^{-1}_{\bb W}\sum_{\sigma_k =\pm1}\int \frac{d\bb s}{(2\pi)^{N_s/2}}\exp\big(-\frac12 \sum_\mu s_\mu^2+\sum_{i\mu} W_{i\mu}\sigma_i s_\mu\big)\Big\vert_{\bb\sigma _{\backslash k}= \bar{\bb\sigma}_{\backslash k}^A}= \\
		&=Z^{-1}_{\bb W}\sum_{\sigma_k =\pm1} \exp\big( \sum_{ij }  J_{ij}(\bb W) \sigma_i \sigma_j   \big)\Big\vert_{\bb\sigma _{\backslash k}= \bar{\bb\sigma}_{\backslash k}^A}=\\&=
		Z^{-1}_{\bb W}\exp \big(J_{kk}(\bb W)+ \sum_{i\neq k,j\neq k}J_{ij }(\bb W)\bar \sigma_i^A\bar \sigma_j^A\big)\sum_{\sigma_k = \pm1 }
		\exp\big(\sum_{i\neq k}J_{ik}(\bb W)\bar \sigma_i ^A\sigma_k\big),
	\end{split}
\end{equation*}
where $J_{ij}(\bb W)= \sum_\mu W_{i\mu}W_{j\mu}$ is the Hebbian kernel associated to $\bb W$. Similarly
\begin{equation*}
	\begin{split}
		P_{\bb W}(\bb\sigma = \bar{\bb\sigma}^A)= Z^{-1}_{\bb W}\exp \big(J_{kk}(\bb W) + \sum_{i\neq k,j\neq k}J_{ij }(\bb W)\bar \sigma_i^A\bar \sigma_j^A\big)
		\exp\big(\sum_{i\neq k}J_{ik}(\bb W)\bar \sigma_i ^A\bar \sigma_k^A\big)
	\end{split}
\end{equation*}
{and, in particular,}
\begin{equation*}
	P (\sigma_k = \bar\sigma _k ^A \vert\bar{ \bb \sigma}_{\backslash k}^A)= \frac{\exp\big(\sum_{i\neq k}J_{ik}(\bb W)\bar \sigma_i ^A\bar \sigma_k^A\big)}{\sum_{\sigma_k = \pm1 }
		\exp\big(\sum_{i\neq k}J_{ik}(\bb W)\bar \sigma_i ^A\sigma_k\big)}.
\end{equation*}
Once we have easily-computable quantities, we take the logarithm of the product over all possible $k$ entries and the whole dataset, thus defining the pseudo-loglikelihood:
\begin{equation}
\begin{split}
	 \mathcal L_{\mathcal D} (\bb W)&= \frac1{N M }\sum_{A=1}^M \sum_{k=1}^N \log \frac{\exp\big(\sum_{i\neq k}J_{ik}(\bb W)\bar \sigma_i ^A\bar \sigma_k^A\big)}{\sum_{x = \pm1 }
		\exp\big(\sum_{i\neq k}J_{ik}(\bb W)\bar \sigma_i ^A x\big)}=\\
  &= \frac1{N M }\sum_{A=1}^M \sum_{k=1}^N \Big(\sum_{i\neq k}J_{ik}(\bb W)\bar \sigma_i ^A\bar \sigma_k^A-\log{\sum_{x = \pm1 }
		\exp\big(\sum_{i\neq k}J_{ik}(\bb W)\bar \sigma_i ^A x\big)}\Big).
\end{split}
\end{equation}

\bibliographystyle{unsrt}
\bibliography{BMs.bib}

\end{document}